\documentclass[12pt]{article}
\usepackage{amsmath}
\usepackage{amssymb}
\usepackage{psfrag,epsf}
\usepackage{enumerate}

\newcommand{\blind}{0}

\usepackage[english]{babel}

\usepackage{bbm}

\usepackage{multirow}
\usepackage{multicol}
\usepackage{makecell} 
\usepackage[thinlines]{easytable}
\usepackage{float}
\usepackage{lscape}
\setcellgapes{10pt}
\usepackage{booktabs}

\usepackage[
backend=biber,
style=numeric,
sorting=none
]{biblatex}
\usepackage{soul} 

\newcommand{\cD}{\mathcal{D}}
\newcommand{\cY}{\mathcal{Y}}
\newcommand{\cX}{\mathcal{X}}
\newcommand{\cN}{\mathcal{N}}

\newcommand{\mR}{\mathbb{R}}

\newcommand{\Y}{\mathbf{Y}}

\newcommand{\X}{\mathbf{X}}

\newcommand{\x}{\mathbf{x}}

\newcommand{\w}{\mathbf{w}}

\newcommand{\T}{\top}

\newcommand{\tA}{{\Tilde{A}}}
\newcommand{\tT}{{\Tilde{T}}}

\newcommand{\lparens}[1]{\left( #1 \right)}

\usepackage{xcolor}

\usepackage[normalem]{ulem} 

\usepackage{kotex}

\usepackage{graphicx}
\usepackage{rotating}
\usepackage{placeins}
\usepackage{capt-of}
\usepackage{caption} 
\usepackage{float}

\begin{document}
\begin{refsection}

\def\spacingset#1{\renewcommand{\baselinestretch}%
{1}\small\normalsize} \spacingset{1}

\if0\blind
{
  \title{Tree-based methods for length-biased survival data}
  \author{
    Jinwoo Lee$^{1,2}$, Hyunwoo Lee$^3$, Donghwan Lee$^{4,*}$, and Jiyu Sun$^{1,*}$ \\\\
    \normalsize Integrated Biostatistics Branch, National Cancer Center$^1$ \\
    \normalsize Department of Biostatistics and Computing, Yonsei University$^2$ \\
    \normalsize Division of Pulmonary and Critical Care Medicine, Department of Internal Medicine, \\
    \normalsize Seoul National University College of Medicine, \\
    \normalsize Seoul Metropolitan Government-Seoul National University Boramae Medical Center$^3$ \\
    \normalsize Department of Statistics, Ewha Womans University$^4$ \\
    \normalsize *Corresponding authors: \texttt{jiyu.sun0@gmail.com}, \texttt{ldh1021@gmail.com}
  }
  \date{}
  \maketitle
} \fi

\if1\blind
{
  \begin{center}
    {\LARGE\bf Title}
  \end{center}
} \fi

\bigskip
\begin{abstract}
Left-truncated survival data commonly arise in prevalent cohort studies, where only individuals who survive from the onset of the disease to enrollment in the study are observed. Such sampling induces length bias, and individuals with longer survival times are more likely to be included in the study. Under a stationary Poisson onset process, the resulting length-biased distribution has a well-established probabilistic characterization. Consequently, conventional methods developed for incident cohort data cannot be applied directly without accounting for the induced sampling bias.
While tree-based methods developed for left-truncated data can be applied, they may be inefficient for length-biased data, as they do not account for the distribution of truncation times. To address this, we propose new survival trees and forests for length-biased right-censored data within the conditional inference framework. Our approach uses a score function derived from the full likelihood to construct permutation test statistics for variable splitting. For survival prediction, we consider two estimators of the unbiased survival function that differ in statistical efficiency and computational complexity. These elements improve efficiency in tree construction and improve the accuracy of survival prediction in ensemble settings. Simulation studies demonstrate efficiency gains in both tree recovery and survival prediction. We further illustrate the utility of the proposed methods using lung cancer data from the Cancer Public Library Database, a nationwide cancer registry in South Korea.
\end{abstract}

\noindent%

\newpage 

\section{Introduction}\label{sec:intro}
Survival analysis is fundamental in epidemiological and clinical research for studying the natural history of disease. 
In epidemiological studies, incident and prevalent cohort designs are two major approaches for investigating time-to-event outcomes. 
Incident cohort studies follow disease-free individuals from disease onset initially to failure; however, for rare diseases or outcomes with long survival times, such designs may require lengthy follow-up periods and large sample sizes to accumulate sufficient events. 
As an alternative, prevalent cohort studies enroll individuals who have already experienced disease onset and follow them prospectively for subsequent outcomes, providing a more practical and efficient sampling design in many settings. 

However, prevalent sampling induces left truncation because only individuals whose failure times exceed their enrollment times are observed. Consequently, individuals with longer survival durations are more likely to be included in the study, leading to a selection bias toward longer survivors \parencite{wang1991nonparametric, wang1993statistical, huang2011nonparametric}. Statistical methods developed for standard right-censored survival data are therefore not directly applicable to such biased samples.
Extensive statistical literature has addressed left-truncated data, including nonparametric one-sample estimators \parencite{turnbull1976empirical, wang1991nonparametric, tsai1987note}, two-sample tests \parencite{lagakos1988nonparametric, pan1998rank, shen2007general}, and regression models such as Cox and accelerated failure time models \parencite{andersen1982cox, wang1993statistical, lai1991rank}.
In addition, a growing body of work has extended tree-based and ensemble methods to accommodate left-truncated right-censored (LTRC) data and time-varying covariates, including \parencite{bacchetti1995survival, wallace2014time, bertolet2016tree, fu2017survival, yao2022ensemble}. 
However, because these methods are developed for general left-truncated settings in which the left-truncation time distribution is unspecified, they may fail to leverage additional information available under a more structured disease-onset process.

Tree-based methods partition the covariate space into homogeneous subgroups through binary splits that effectively capture complex covariate interactions. 
The most widely used algorithm is CART, which exhaustively searches all covariates and cutpoints to minimize an impurity measure. 
However, CART is prone to selection bias toward variables with many possible splits \parencite{kass1980exploratory, segal1988regression} or missing values \parencite{kim2001classification}, and provides no formal test of split significance. Although pruning can control tree size, its inherent bias can still limit interpretability.
Conditional inference trees (CIT; \parencite{hothorn2006unbiased}) address this issue by testing the global null of independence via permutation tests, which ensures an unbiased choice of splitting variable. 
\textcite{fu2017survival} adapted both CART and CIT to LTRC data--hereafter referred to as LTRC-CART and LTRC-CIT--and showed that LTRC-CIT tends to perform better than LTRC-CART in recovering true tree structure and in survival prediction, particularly under small sample sizes or high censoring. 
More recently, \textcite{yao2022ensemble} extended LTRC-CIT into an ensemble framework (LTRC-CIF) by growing bootstrap forests of LTRC-CITs and introducing an out-of-bag tuning procedure to regulate tree complexity. 
Nonetheless, both LTRC-CIT and LTRC-CIF compute split statistics and survival estimates from the conditional likelihood given truncation. 
As a result, they fall short of the efficiency achievable under full-likelihood alternatives.

Under the assumption that disease onset follows a stationary Poisson process, often referred to as the stable disease assumption, prevalent sampling induces a special form of left truncation known as length-biased sampling \parencite{wang1991nonparametric, huang2011nonparametric}. This sampling structure yields closed-form expressions for the joint and marginal distributions of the truncation time (from disease onset to enrollment) and residual survival time (from enrollment to failure), as well as their symmetry properties \parencite{lancaster1990econometric}.
Leveraging these properties, a variety of efficient estimators have been developed for length-biased right-censored (LBRC) data using various statistical techniques. 
\textcite{vardi1989multiplicative} established the nonparametric maximum likelihood estimator (NPMLE) for the one-sample survival function based on full likelihood. 
\textcite{ning2010non} derived a nonparametric score test statistic from the full likelihood under proportional hazards alternatives. 
For the Cox model, \textcite{qin2010statistical} introduced a weighted estimating equation approach, while \textcite{qin2011maximum} developed an expectation-maximization (EM) algorithm to estimate both the regression coefficients and the baseline hazard function via full likelihood. 
\textcite{huang2012composite} proposed a composite partial likelihood of the Cox model that effectively doubles the information from the truncation time by incorporating residual survival time. 
For other approaches that aim to improve efficiency in regression models, see the review by \textcite{shen2017nonparametric_review}. 
More recent work has extended inference to the estimation of quantile residual lifetime \parencite{wang2017nonparametric} and restricted mean survival time under LBRC settings \parencite{lee2018analysis, he2020nonparametric, bai2024inference}. 
To date, however, no tree-based or ensemble learning methods have been specifically developed for LBRC data.

To fill this gap, we introduce novel survival trees and forests for LBRC settings. 
Specifically, we embed a score function based on the full likelihood within the CIT framework and use existing efficient, nonparametric, unbiased survival estimators for prediction in CIT and CIF. 
We implement two variants to balance statistical efficiency and computational scalability by incorporating existing approaches: a nonparametric estimator that maximizes the full likelihood \parencite{vardi1989multiplicative} and a closed-form nonparametric estimator based on the composite conditional likelihood \parencite{wang2017nonparametric, he2020nonparametric}. 
The remainder of this paper is organized as follows. 
In Section~\ref{sec:methods}, we review the structures of length-biased data, summarize conditional inference trees and forests for left-truncated data, and present our adaptations for the LBRC data. 
In Section~\ref{sec:simulation}, we study the properties of the proposed tree methods and compare them to existing methods for LTRC data. 
In Section~\ref{sec:application}, we apply the proposed methods to lung cancer data from the Cancer Public Library Database, a nationwide cancer registry in South Korea, where we conduct stationarity assessment and predictive evaluation using cross-validated Brier scores.  
We conclude with final remarks and future directions in Section~\ref{sec:discussion}.

\section{Tree-based methods for LBRC data} \label{sec:methods}
\subsection{Length-biased right-censored data}
\begin{figure}[t]
    \centering
    \includegraphics[width=0.7\textwidth]{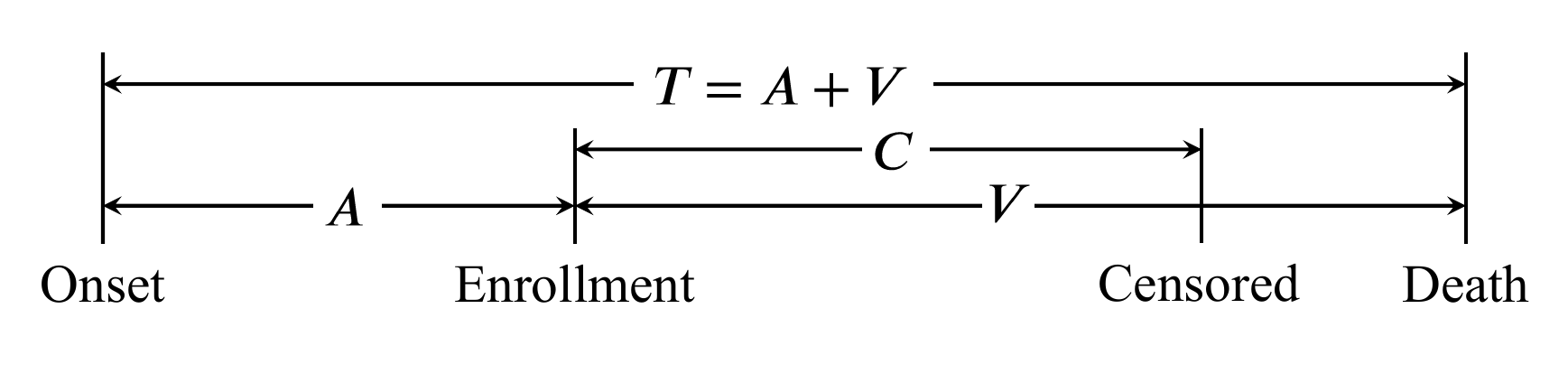}
    \caption{Illustration of the length-biased sampling structure.}
    \label{fig1:LB_sampling}
\end{figure}

Consider a prevalent cohort study in which patients are enrolled after disease onset and followed prospectively until failure or the end of the study.
Let $\tT$ denote the duration from disease onset to failure in the incident population, with unbiased density function $f(t) = dF(t)/dt$ and survival function $S(t)$, and let $\tA$ denote the duration from onset to study enrollment in the same population.
Both $\tT$ and $\tA$ are defined at the population level and are not directly observed. Instead, the prevalent cohort consists only of individuals satisfying $\tilde A<\tilde T$, leading to an over-representation of individuals with longer failure times.
Under the stable disease assumption, in which disease onset follows a stationary Poisson process, the enrollment time relative to disease onset is uniformly distributed over the disease duration. For a sampled patient, we observe the truncation time $A$ (the backward recurrence time, from onset to enrollment) and the residual survival time $V$ (the forward recurrence time, from enrollment to failure), and the corresponding observed failure time is $T = A + V$, which follows the length-biased distribution induced by the sampling mechanism.
Figure~\ref{fig1:LB_sampling} illustrates this sampling structure.

Let \( f_T(t), f_A(t) \), and \( f_V(t) \) denote the marginal density functions of \( T, A \), and \( V \), respectively;
\( F_T(t), F_A(t) \), and \( F_V(t) \) their corresponding distribution functions; and \( S_T(t), S_A(t) \), and \( S_V(t) \) their corresponding survival functions.
Following the renewal theory in Lancaster~\autocite[Ch.~3]{lancaster1990econometric}, the joint density of \( (A, T) \) under length-biased sampling is
\begin{equation}\label{eq:joint_AT}
f_{A,T}(a, t) = \frac{f(t)}{\mu} I(t \geq a > 0),
\end{equation}
where \( \mu = \int_0^\infty u f(u)\, du \) is the mean of \( \tT \). The marginal density functions are then given by
\begin{align}
f_T(t) &= \frac{t f(t)}{\mu} I(t > 0), \label{eq:fT}\\
f_A(t) = f_V(t) &= \frac{S(t)}{\mu} I(t > 0). \label{eq:fAV}
\end{align}

When \( V \) is subject to right-censoring, we have the event time \( Z = \min(T, A + C) \) and \( O = \min(V, C) \), where \( C \) denotes the time from enrollment to censoring.
The censoring indicator is defined as \( \delta = I(T \leq A + C) \), and we assume that \( (A, V) \) is independent of \( C \).
The observed data thus consist of \( \{(Z_i, A_i, \delta_i)\}_{i=1}^{n} \) for \( n \) independent subjects.
This constitutes a special case of LTRC data, referred to as LBRC data.
Under the LBRC setting, by treating the censoring mechanism as a nuisance component, the full likelihood is proportional to
\begin{equation}\label{eq:full_lik}
\mathcal{L} = \prod_{i=1}^{n} \frac{f(Z_i)^{\delta_i} S(Z_i)^{1-\delta_i}}{\mu}
\end{equation}
and can be decomposed into
\begin{equation}\label{eq:lik_decomp}
\mathcal{L} = \left\{ \prod_{i=1}^{n} \frac{f(Z_i)^{\delta_i} S(Z_i)^{1-\delta_i}}{S(A_i)} \right\} \left\{ \prod_{i=1}^{n} \frac{S(A_i)}{\mu} \right\} = \mathcal{L}_C \times \mathcal{L}_M,
\end{equation}
where \( \mathcal{L}_C \) is the conditional likelihood of the observed failure time \( Z \) given the truncation time \( A \), and \( \mathcal{L}_M \) is the marginal likelihood of \( A \).

Next, let \( \mathbf{X} \in \mathcal{X} = \prod_{j=1}^{m} \mathcal{X}_j \) denote the vector of \( m \)-dimensional baseline covariates.
It is commonly assumed that, conditional on \( \mathbf{X} \), the censoring time \( C \) is independent of \( (A, V) \).
With observed data \( \{(Z_i, A_i, \delta_i, \mathbf{X}_i)\}_{i=1}^{n} \), we now proceed to describe conditional inference trees and forests for LTRC survival data, which aim to estimate the unbiased survival function \( S(t \mid \mathbf{X}) \).

\subsection{Conditional inference trees and forests for LTRC data}
Conditional inference trees are built using the recursive partitioning method of \textcite{hothorn2006unbiased}. 
At each node, an unbiased variable selection procedure is performed: a permutation-based test assesses the global null hypothesis of independence between the response variable \(\Y \in \mathcal{Y}\) and the covariates in \(\X \in \cX \).
If this null is rejected, the covariate with the strongest association to \(\Y\) is selected, and the optimal binary split within that covariate is determined. 
The recursion stops when the null hypothesis cannot be rejected at the specified significance level. 

Within a CIT, the association between \(X_j\) and \(\Y\) is evaluated based on a random sample \(\cD_n = \{ (\Y_i, \X_{i})\}_{i=1}^n\), using linear statistics of the form
\begin{equation}\label{eq:linear_stat}
    T_j(\cD_n, \w) = vec\lparens{\sum_{i=1}^n w_i g_j(X_{ji}) h(\Y_i,(\Y_1,...,\Y_n))^\T} \in \mR^{p_j q}, 
\end{equation}
where $\w:= (w_1,...,w_n)$ is a vector of non-negative integer-valued case weights having nonzero elements when the corresponding observations are elements of the node and zero otherwise, $g_j: \mathcal{X}_j \rightarrow \mR^{p_j}$ is a nonrandom transformation of covariate $X_j$, and $h: \cY \times \cY^n \rightarrow \mR^q$ is the influence function and depends on the responses $(\Y_1,...,\Y_n)$ in a permutation-symmetric way. 
Then the permutation-based test of the global null hypothesis of independence is performed on univariate test statistics, which consist of standardized \(T_j(\cD_n,\w)\) using the conditional expectation $\mu_j$ and covariance $\Sigma_j$ \parencite{strasser1999asymptotic}.

The log-rank score is typically used as the influence function for right-censored data \parencite{hothorn2006unbiased}, and \textcite{fu2017survival} adopted the extended log-rank score of \textcite{pan1998rank} for the LTRC setting.

Given the \(i\)th response variable \(\Y_i =(Z_i, A_i, \delta_i)\), the influence function is defined as
\begin{equation}\label{eq:logrank_lt}
    h(\Y_i,(\Y_1,...,\Y_n)) = U^{LT}_i = \delta_i + \log  \hat  S(Z_i) - \log \hat S(A_i),
\end{equation}
where $\hat{S}(t)$ is the estimator for the unbiased survival function $S(t)$. 
The utility of the log-rank score in the CIT algorithm is to assign the univariate scalar value $U_i^{LT}$ to the multivariate response \(\Y_i\), so the algorithm can execute in the same way as in the case of the univariate numeric response. 
For estimating the unbiased survival function, \textcite{fu2017survival} uses the Kaplan-Meier estimator with an adjusted risk set
\begin{equation}\label{eq:LT}
   \hat S (t) = \prod_{u \in [0,t]} \left\{ 1-  \frac{\sum_{i=1}^n \delta_iI(Z_i = u)}{\sum_{i=1}^n  I(A_i \leq u \leq Z_i)}\right\},
\end{equation}
which corresponds to the NPMLE of the conditional likelihood \(\mathcal{L}_C\) without specifying any distribution for \(A\) \parencite{wang1991nonparametric}.

Conditional inference forests (CIFs) are an ensemble of CITs that employ a prediction procedure based on an adaptive local weighting scheme \parencite{hothorn2004bagging, meinshausen2006quantile, athey2019generalized}. 
Suppose that a set of $B$ trees are grown. For any new data \(\x \in \cX\), let \(\cN_b(\x)\) denote the set of observed data \(\{\X_i\}_{i=1}^n\) that reside in the same terminal node as \(\x\) in the $b$-th tree. 
The "nearest neighbor" forest weight $\alpha_{i}(\x)$ is computed as
\begin{equation}\label{eq:forest_weights}
    \alpha_{i}(\x) = \frac{1}{B}\sum_{b=1}^B \frac{I(\X_i \in \cN_b(\x))}{\sum_{j=1}^n I(\X_j \in \cN_b(\x))},
\end{equation}
which sum to one. 
The weight \(\alpha_i(\x)\) measures the "similarity" of the observation \(\X_i\) in the original sample across $B$ trees in the forest.
The unbiased survival prediction for the new data $\x$, i.e., \(S(t|\x)\), is then obtained by incorporating these weights into the KM estimator in \eqref{eq:LT}. Note that prediction with a single CIT can also be viewed through the forest-weight framework, which, in this case, corresponds to the Kaplan–Meier estimate at the terminal node.

The influence function \eqref{eq:logrank_lt} and the estimator of the unbiased survival function \eqref{eq:LT} can be further refined for the LBRC setting, which we describe next.

\subsection{CIT and CIF for LBRC data}
The log-rank score for general LTRC data in \eqref{eq:logrank_lt} is derived from the conditional likelihood $\mathcal{L}_C$ given the individual truncation time $A_i$. While this avoids modeling the truncation distribution, it does not utilize the additional structure available under length-biased sampling. In contrast, the truncation mechanism can be explicitly modeled through the marginal likelihood $\mathcal{L}_M$, leading to the score function proposed by \textcite{ning2010non}:
\begin{equation} \label{eq:logrank_lb}
    U_i^{LB} = \delta_i + \log \hat S (Z_i) - \frac{\int_0^\infty  \hat S (t) \log \hat S (t) dt}{\int_0^\infty  \hat S(t)dt}.
\end{equation}
The last term no longer depends on $A_i$ and instead reflects the population-level sampling structure induced by length-biased sampling.
For two-sample testing under proportional hazards, \textcite{ning2010non} showed that test statistics based on \eqref{eq:logrank_lb} achieve greater power for LBRC data than log-rank tests with truncation-adjusted risk sets. Accordingly, we adopt the score in \eqref{eq:logrank_lb} as the influence function within the CIT/CIF framework for variable selection. 
In practice, because the final term is constant across observations, it cancels upon standardization of the linear statistic~\eqref{eq:linear_stat}, so the score reduces to $\delta_i + \log \hat{S}(Z_i)$.

An appropriate estimator of the unbiased survival function is required for both influence function \eqref{eq:logrank_lb} and survival prediction with weights \eqref{eq:forest_weights}.
While the KM estimator in \eqref{eq:LT} can be used,  it ignores the information on \(A\) contained in the marginal likelihood \(\mathcal{L}_M\). 
 To address this, we consider two existing nonparametric estimators for the unbiased survival function in LBRC data, the maximum full-likelihood estimator (MFLE; \parencite{vardi1989multiplicative}) and the maximum composite conditional-likelihood estimator (MCLE; \parencite{wang2017nonparametric, he2020nonparametric}). We detail these approaches in Sections ~\ref{sec:full-likelihood} and ~\ref{sec:composite-likelihood}, respectively. 

We refer to the conditional inference tree and forest methods that incorporate the LBRC influence function~\eqref{eq:logrank_lb} and one of the two unbiased survival estimators (MFLE or MCLE) as LBRC-CIT and LBRC-CIF, respectively. Since either estimator can be used independently for tree construction (i.e., variable selection and splitting with the influence function) and for survival prediction, four variants arise, as summarized in Table~\ref{tab:lbrc_cit_variants}. Their properties and predictive performance are examined in Section~\ref{sec:simulation}.

\begin{table}[h]
\centering
\caption{The four variants of LBRC-CIT, defined by the estimator used for tree construction and survival prediction.}
\label{tab:lbrc_cit_variants}
\begin{tabular}{l|cc}
\toprule
 & \multicolumn{2}{c}{Survival prediction} \\
Tree construction & MFLE (F) & MCLE (C) \\
\midrule
MFLE (F) & LBRC-CIT(F,F) & LBRC-CIT(F,C) \\
MCLE (C) & LBRC-CIT(C,F) & LBRC-CIT(C,C) \\
\bottomrule
\end{tabular}
\end{table}

\subsubsection{Method 1: Maximum full-likelihood estimator (MFLE)} \label{sec:full-likelihood}

\textcite{vardi1989multiplicative} showed that the nonparametric maximum likelihood estimator (NPMLE) of the unbiased survival function $S(t)$ can be obtained by maximizing the full likelihood~\eqref{eq:full_lik} with respect to the distribution of the observed failure times $F_T(t)$:

\begin{equation}
     \prod_{i=1}^ndF_T(Z_i)^{\delta_i} \left\{ \int_{u \geq Z_i }u^{-1} dF_T(u) \right\}^{1-\delta_i},
\end{equation}
followed by a transformation to an unbiased survival function using \eqref{eq:fT}. Specifically, let $0 < t_1 < ... < t_h$ denote the distinct ordered observed times in $\{Z_{i}\}_{i=1}^n$. Define \(q_j := d F_T(t_j)\), subject to \(\sum_{j=1}^h q_j=1\).
The NPMLE for \(dF_T(t_j)\) can be computed via the expectation maximization (EM) algorithm with the following limiting equation:
\begin{equation}\label{eq:V1_2}
    \hat q_j^{(new)} = \frac{1}{n} \sum_{i=1}^n \left\{ \delta_i I(Z_i = t_j) + (1-\delta_i)  \frac{t_j^{-1}\hat q_j^{(old)} I(Z_i\leq t_j)}{\sum_{k=1}^h t_k^{-1} \hat q_k^{(old)} I(Z_i\leq t_k)} \right\}.
\end{equation} 
and \(S(t)\) can be estimated with the transformation
\begin{equation}\label{eq:V1_1}
    \hat S^{LB1}(t_j) = 1 - \frac{\int_0^{t_j} u^{-1}d\hat F_T(u)}{\int_0^{\infty} u^{-1}d\hat F_T(u)} = \sum_{i=j}^h \frac{t^{-1}_i \hat q_i}{\sum_{k=1}^h t^{-1}_k \hat q_k}.
\end{equation}
For prediction, the unbiased survival function for new data is estimated using the weighted version of \eqref{eq:V1_1} with the forest weights \eqref{eq:forest_weights}. For the LBRC influence function \eqref{eq:logrank_lb}, a direct plug-in of \eqref{eq:V1_1} is possible. However, it can lead to numerical instability during the permutation tests used for unbiased variable selection. This instability arises because the constraint \(\sum_{j=1}^h q_j=1\) implies \(\hat S^{LB1}(t_h)=0\), so that \(\log \hat S^{LB1}(t_h)\) = $-\infty$. Even if we replace \(\hat S^{LB1}(t_h)\) with a small constant \(\epsilon\), the resulting value can be substantially different from neighboring estimates at \(t_1, \ldots, t_{h-1}\), potentially reducing the test power. To mitigate this issue, we instead use a discretized version of the hazard function, \(\Lambda(u) = \sum_{u \ge t_j} \lambda_j\), where $\lambda_j$ is the positive jump at $t_j$, estimated as
\begin{equation}
    \hat\lambda_j^{LB1} = \frac{t_j^{-1} \hat q_j}{\sum_{k=j}^h t_k^{-1} \hat q_k}.
\end{equation}
This yields the estimator \(\log \hat S^{LB1}(t) = -\sum_{t \ge t_j} \hat \lambda^{LB1}_j\), which avoids extreme values at time \(t_h\)
. The derivation of this estimator is provided in the supplemental material.

A key advantage of the NPMLE is that, unlike the KM estimator \eqref{eq:logrank_lt}, it additionally leverages information from the marginal likelihood of \(A\), leading to improved efficiency in both variable selection and survival prediction.
Nonetheless, the NPMLE has no closed form and requires an iterative EM algorithm, which can become computationally burdensome when numerous node splits are needed or when the test set is large--- particularly in forests, where the weights vary across test observations. This computational cost motivates the alternative estimator introduced in Section~\ref{sec:composite-likelihood}.

\subsubsection{Method 2: Maximum composite conditional-likelihood estimator (MCLE)} \label{sec:composite-likelihood}
The composite likelihood method \parencite{arnold1988bivariate} was first applied to the semiparametric Cox model by \textcite{huang2012composite}, and later utilized in quantile residual lifetime models \parencite{wang2017nonparametric} and restricted mean survival time estimation \parencite{he2020nonparametric}.
Leveraging the symmetry between truncation time \(A\) and residual time \(V\) from \eqref{eq:fAV}, the density of \(T\) given \(A\) equals to the density of \(T\) given \(V\), and the non-censored data can be treated as generated with truncation time not only \(A\) but also \(V\) \parencite{huang2012composite}. Motivated by this, the composite conditional-likelihood can be constructed as \parencite{he2020nonparametric}
\begin{align}
    \prod_{i=1}^n \bigg\{ \frac{f(Z_i)}{S(A_i)} \times \frac{f(Z_i)}{S(O_i)} \bigg\}^{\delta_i}   \bigg\{ \frac{S(Z_i)}{S(A_i)} \bigg\}^{1-\delta_i} \notag
    = \prod_{i=1}^n \frac{\lambda(Z_i)^{2\delta_i} \exp\left\{ -(1+\delta_i) \Lambda(Z_i) \right\}}{\exp\left[-\left\{ \Lambda(A_i) + \delta_i \Lambda(O_i) \right\}\right]}.
\end{align}
Maximizing this with respect to \(\Lambda(t)\) yields a closed-form estimator
\begin{equation}\label{eq:ccl_S}
    \hat S^{LB2}(t) = \prod_{u \in [0,t]} \left\{ 1- d\hat\Lambda^{LB2}(u)\right\},
\end{equation}
where
\begin{equation}\label{eq:ccl_CHF}
    \hat \Lambda^{LB2}(t) = \sum_{i=1}^n \frac{2\delta_i I(Z_i \leq t)}{\sum_{j=1}^n \left\{ I(A_j \leq t \leq Z_j) + \delta_j I(O_j \leq t \leq Z_j) \right\}}.
\end{equation}
For the LBRC influence function \eqref{eq:logrank_lb}, either \(\hat S^{LB2}(t)\) can be used with a log-transformation, or \(\hat \Lambda^{LB2}(t)\) can be substituted via the relation \(\Lambda(t) = -\log S(t)\); we adopt the latter for numerical stability. For unbiased survival prediction of new data, the weighted version of \eqref{eq:ccl_S} is used.

Intuitively, this approach improves efficiency by combining information from $A$ and $V$, compared to (\ref{eq:LT}), which uses only information from $A$. In addition, the closed-form nature facilitates faster recursive partitioning and survival prediction than the iterative NPMLE. However, the performance of the variable selection and survival prediction based on this estimator may be inferior to that of the estimator in Section~\ref{sec:full-likelihood}, which maximizes the full-likelihood. 

\subsection{Regulating the construction of the trees in the proposed CIFs}
Tree complexity in conditional inference forests is controlled by several tuning parameters: \texttt{mtry} (the number of covariates randomly selected at each split), \texttt{minsplit} (the minimum sum of case-weights required to initiate a split), \texttt{minprob} (minimum proportion of observations needed to form a terminal node), and \texttt{minbucket} (the minimum sum of case-weights allowed in a terminal node). 
While the default values---\(\texttt{mtry} = \sqrt{p}\) \parencite{hothorn2006survival}; \(\texttt{minsplit}=20\), \(\texttt{minbucket}=7\) in the \texttt{cforest} implementation---are commonly used, prior work has shown that performance is sensitive to these parameters. In particular, \textcite{yao2021ensemble, yao2022ensemble} demonstrate that careful tuning improves the predictive accuracy of conditional inference forests for both interval-censored and left-truncated right-censored data.
To address this, they propose selecting \texttt{mtry} by minimizing the integrated Brier score (IBS; \cite{graf1999assessment}) using out-of-bag (OOB) observations, demonstrating improved and stable performance across scenarios compared to default settings.

As length-biased right-censored data are a special case of LTRC data, we build on the tuning strategy proposed by \textcite{yao2022ensemble}. For the \(b\)th tree, OOB observations are those excluded from its bootstrap sample and thus not used during the corresponding \(b\)th tree construction. 
The survival probability for observation \(\x_i\) is estimated using only the subset of \(B\) trees in which \(\x_i\) is OOB, denoted as \(\hat S^{OOB}_{\x_i}(t)\). The prediction target \(\hat S_{\x}(t)\) in \textcite{yao2022ensemble} for LTRC data is defined as
\begin{equation} \label{eq:target_conditional}
    \hat S_{\x_i} (t) = \begin{cases}
        1, & t \in [0,A_i); \\
        \hat S(t|\x_i)/\hat S(A_i|\x_i), & t \in [A_i,\max(Z_i)),
    \end{cases}
\end{equation}
where $\hat S (t | \x)$ is an estimated unbiased survival function for $\x$. This formulation adjusts for the delayed entry of each training subject.

To evaluate the fit of the CIF for a given \texttt{mtry}, the estimation error IBS is computed, where the Brier score at time \(t\) is defined as
\begin{equation}
\widehat{\mathrm{BS}}(t;\hat S, \mathcal{D}_n) = \frac{1}{|\mathcal{D}_n|} \sum_{i=1}^n 
\widehat W_i(t) \left\{ I(Z_i > t) - \hat S^{OOB}_{\x_i} (t) \right\}^2,  
\end{equation}
and the corresponding IBS is
\begin{equation}
\widehat{\mathrm{IBS}}(\hat S; \mathcal{D}_n) = \frac{1}{\max(Z_i)}  \int_0^{\max(Z_i)} 
\widehat{\mathrm{BS}}(t;\hat S, \mathcal{D}_n) dt.
\end{equation}
Each \(\widehat W_i(t)\) is an inverse probability of censoring weights given by 
\begin{equation}
\widehat W_i(t) = \frac{(1 - I(Z_i > t)) \delta_i}{\hat G(Z_i)} + \frac{I(Z_i > t)}{\hat G(t)},
\end{equation}
with \(\hat G\) the Kaplan-Meier estimator of the censoring distribution. 

As an alternative tuning criterion, we also considered the out-of-bag concordance index (C-index), selecting the \texttt{mtry} value that maximized predictive concordance among comparable observation pairs using OOB predicted cumulative risk scores. Detailed definitions of the risk score used in the C-index are provided in the Supplementary Material.

For the remaining tuning parameters, \textcite{yao2022ensemble} recommends adapting \texttt{minsplit} and \texttt{minbucket} to the sample size by setting both to \(\max(\text{default}, \sqrt{n})\). We adopt the same strategy for our LBRC data analysis. For further details on tuning procedures, see \textcite{yao2021ensemble, yao2022ensemble}.

\section{Simulation Study} \label{sec:simulation}
In this section, we conduct a series of simulations to study the properties of the proposed LBRC-CITs and LBRC-CIFs. For LBRC-CITs, we assess (i) the unbiasedness of our LBRC-CIT split selection, (ii) the ability to recover the true tree structure, and (iii) prediction accuracy with respect to the unbiased survival function relative to the benchmark LTRC-CIT. For LBRC-CIFs, we evaluate (i) the impact of \texttt{mtry} tuning and (ii) prediction performance under realistic test settings relative to LTRC-CIF, LTRC-CIT, LBRC-CITs, and two Cox models: one for left-truncated data and a more efficient version for length-biased data.

Throughout all simulation studies, LBRC data are generated as follows: we first generate the unbiased failure time \(\tilde{T}\) under the true distribution, then draw the onset time \(\tilde{A} \sim U(0, \tau)\), where \(\tau\) is chosen to exceed the upper bound of \(\tilde{T}\) to satisfy the stationarity condition \parencite{addona2006formal, shen2009analyzing}. We retain only observations where \(\tilde{T} > \tilde{A}\), yielding the observed failure time \(T = A + V\). The censoring time, measured from enrollment $C$, is generated from an exponential distribution for various censoring percentages. The event indicator is defined as \(\delta = \mathbb{I}(T \leq A + C)\).

\subsection{Properties of the LBRC-CITs} 
We evaluate the proposed LBRC-CITs, in which both tree construction and survival prediction can be carried out using either MFLE or MCLE. To disentangle the impact of the estimator used for each task, we consider the four variants defined in Table~\ref{tab:lbrc_cit_variants}.

We first investigate the unbiasedness of variable selection and the ability to recover the correct tree structure using LBRC-CIT(F,$\cdot$) and LBRC-CIT(C,$\cdot$). We then assess all four variants in terms of prediction accuracy, comparing their performance against the benchmark LTRC-CIT method~\parencite{fu2017survival} to evaluate potential efficiency gains.

\subsubsection{Test of unbiased variable selection} \label{unbiased}
The CIT algorithm \parencite{hothorn2006unbiased} is known to be unbiased in selecting splitting variables under the null, i.e., when the response is independent of all covariates, each variable is selected with equal probability. Since the LBRC-CIT inherits this selection procedure, it is expected to maintain the same property. Simulation results confirm this expectation. Full details of the testing procedure and the results are provided in the supplementary material.

\subsubsection{Recovering the correct tree structure} \label{sec:rc_rate}
We next examine the ability of the proposed methods to recover the underlying tree structure generating the data. 
The LTRC-CIT method \parencite{fu2017survival} is included as a benchmark for comparison. 
The simulation setup is as follows:

We generate 30 covariates $X_1,...,X_{30}$, where for each $j=0,...,9$, the covariate \(X_{3j+1}\) is a categorical variable taking values randomly in the set $\{1,2,3,4,5,6\}$, \(X_{3j+2}\) is binary taking values randomly from $\{0,1\}$, and \(X_{3j+3}\) is drawn from uniform distribution $U(0,1)$. 
Only the first three covariates, \( X_1, X_2, \) and \( X_3 \), determine the unbiased survival function of the failure time \( \tT \). 

The unbiased failure time \( \tT \) follows one of four distributions, \( \tT_1, \tT_2, \tT_3, \) or \( \tT_4 \) defined as follows:
\begin{itemize}
    \item When $X_1\leq 3$
    \begin{itemize}
        \item if $X_2\leq1$, distributed as $\tT_1$
        \item if $X_2>1$, distributed as $\tT_2$
    \end{itemize}
    \item When $X_1> 3$
    \begin{itemize}
        \item if $X_3\leq1/2$, distributed as $\tT_3$
        \item if $X_3>1/2$, distributed as $\tT_4$,
    \end{itemize}
\end{itemize}
\noindent where each \(\tilde{T}_i\) corresponds to a specific distribution with distinct parameters:

\begin{itemize}

    \item Weibull increasing (WI) hazard: shape \(\alpha = 2.0\), scale \(\beta \in \{2.0, 3.5, 6.0, 10.0\}\);
    \item Weibull decreasing (WD) hazard: shape \(\alpha = 0.9\), scale \(\beta \in \{7.0, 3.0, 2.5, 1.0\}\);
    \item Lognormal (Lgn): \((\mu, \sigma) \in \{(2.0, 0.3), (1.8, 0.2), (1.2, 0.3), (0.5, 0.5)\}\);
    \item Bathtub-shaped hazard (Bat, \parencite{hjorth1980reliability}): survival function
    \begin{equation*}
        S(t;a,b,c) = \frac{\exp\left(-\frac{1}{2}at^2\right)}{(1+ct)^{b/c}}
    \end{equation*}
    with \(b = 1, c = 5, a \in \{0.01, 0.06, 0.20, 0.70\}\).
\end{itemize}

Within each distribution, the four parameter settings listed above are assigned, in order, to the four leaves $\tilde{T}_1, \tilde{T}_2, \tilde{T}_3, \tilde{T}_4$ of the true tree, and the simulation is repeated across all four distributions (WI, WD, Lgn, and Bat).
To see the impact of censoring rate and sample size, we considered two censoring rates \(c \in \{0.2, 0.5\}\) and three sample sizes \(n \in \{100, 200, 400\}\), resulting in 6 simulation settings per distribution. 
For each setting, we conducted 1000 simulation replicates to evaluate the performance of LBRC-CIT(F,\(\cdot\)) and LBRC-CIT(C,\(\cdot\)), and to compare them with the LTRC-CIT method.
The recovery rate is defined as the proportion of simulation replicates in which the selected covariates and their corresponding split structure exactly match the true tree used to generate the data.

\begin{figure}[t]
    \centering
    \includegraphics[width=0.95\textwidth]{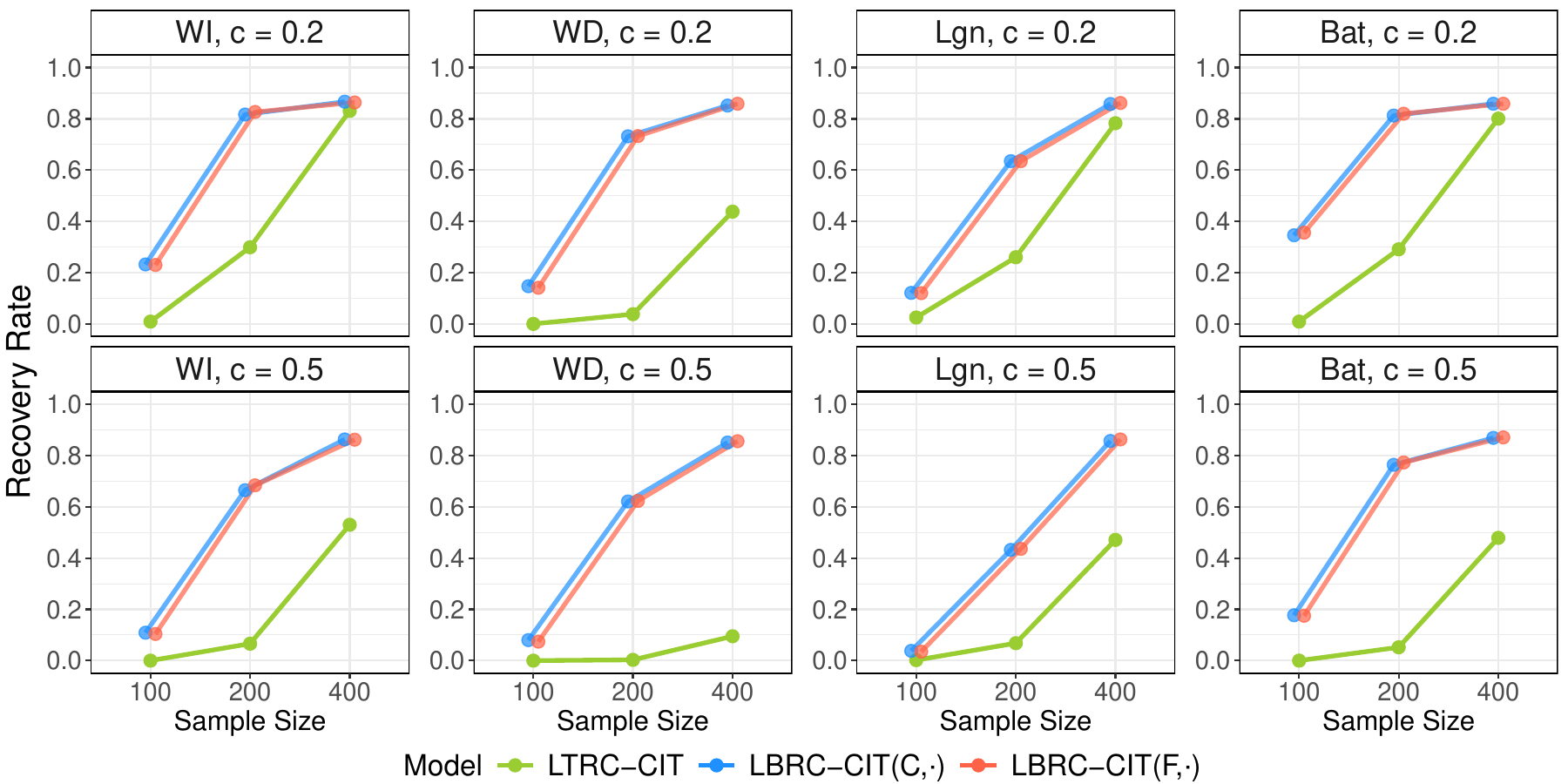}
    \caption{Tree structure recovery rates for LTRC-CIT, LBRC-CIT(C,\(\cdot\)), and LBRC-CIT(F,\(\cdot\)) under varying sample sizes, censoring rates, and survival distributions. Higher values indicate more accurate recovery of the true underlying tree structure.}
    \label{fig2:Recovery_Rate}
\end{figure}

Figure~\ref{fig2:Recovery_Rate} shows that both LBRC-CIT(F,$\cdot$) and LBRC-CIT(C,$\cdot$) consistently outperform LTRC-CIT across all settings. Although higher censoring rates reduce overall recovery performance, the improvement offered by the LBRC-CITs becomes increasingly pronounced as the sample size increases from $n=100$ to $200$ and $400$. Notably, for the Weibull distribution with decreasing hazard and a 50\% censoring rate, the recovery rate of LTRC-CIT remains around 10\% even at a sample size of 400, whereas LBRC-CITs achieve recovery rates exceeding 80\%. These results suggest that LBRC-CITs enable more efficient tree reconstruction and exhibit greater robustness to heavy censoring than LTRC-CIT.

When comparing LBRC-CIT(F,$\cdot$) and LBRC-CIT(C,$\cdot$), the difference in recovery performance is negligible. This indicates that using MCLE for tree construction retains nearly the same efficiency as MFLE, even under challenging conditions such as small sample sizes ($n=100$) or high censoring rates ($c=0.5$) across all distribution scenarios.

\subsubsection{Recovering the unbiased survival function}\label{sec:pred_acc}

Using a factorial design, we compare the accuracy of estimating the unbiased survival function of all four LBRC-CIT variants against LTRC-CIT across combinations of failure time distribution, sample size, censoring rate, tree construction method, and survival prediction method.
We generate 30 covariates as in Section~\ref{sec:rc_rate} and consider four data-generating structures, each defined by how the covariates $X_1, X_2, X_3$ determine the failure time $\tT$:
\begin{itemize}
  \item \textbf{Tree:} the tree-structured data of Section~\ref{sec:rc_rate};
  \item \textbf{Linear:} $\vartheta = \phi_0 + \sum_{j=1}^{3}\phi_j X_j$;
  \item \textbf{Nonlinear:} $\vartheta = \phi_0 + \phi_1 \cos(\pi(X_1 + X_2)) + \phi_2 \sqrt{X_1 + X_2} + \phi_3 X_3^{X_2}$;
  \item \textbf{Interaction:} $\vartheta = \phi_0 + \phi_1 X_1 X_2 X_3 + \phi_2 X_3^3$,
\end{itemize}
where $\vartheta$ is a location parameter.
For the Tree structure, the four distributions of Section~\ref{sec:rc_rate} are reused.
For the Linear, Nonlinear, and Interaction structures, the failure time $\tT$ is generated from a Weibull distribution with either a decreasing hazard (shape $\alpha = 0.8$) or an increasing hazard (shape $\alpha = 2.0$), with scale $\beta = e^{-\vartheta}$ and the location parameters given in Table~\ref{tab:loc_params}.

\begin{table}[h]
\centering
\caption{Location parameters $\phi_1, \phi_2, \phi_3$ and intercept $\phi_0$ used to generate $\vartheta$ under the Linear, Nonlinear, and Interaction structures. The intercept $\phi_0$ is given separately for the decreasing- and increasing-hazard Weibull distributions.}
\label{tab:loc_params}
\begin{tabular}{lccccc}
\toprule
 & & & & \multicolumn{2}{c}{$\phi_0$} \\
\cmidrule(lr){5-6}
Structure & $\phi_1$ & $\phi_2$ & $\phi_3$ & Decreasing & Increasing \\
\midrule
Linear      & $1$            & $1$             & $-\tfrac{1}{3}$ & $-\log 1$ & $-\log 2$ \\
Nonlinear   & $1$            & $1$             & $\tfrac{1}{6}$  & $-\log 5$ & $-\log 10$ \\
Interaction & $\tfrac{1}{2}$ & $-\tfrac{3}{2}$ & --              & $-\log 1$ & $-\log 2$ \\
\bottomrule
\end{tabular}
\end{table}

For each failure time distribution under each structure, we use the same six settings ($c \in \{0.2, 0.5\}$, $n \in \{100, 200, 400\}$) as in Section~\ref{sec:rc_rate}, with 500 simulation replicates each.
Estimation accuracy is quantified by the average integrated $L_2$-distance between the target and estimated survival curves:
\begin{equation}\label{eq:L2}
L_2(\hat{S}^{\dagger}) = \frac{1}{n}\sum_{i=1}^{n} \frac{1}{\eta_i} \int_0^{\eta_i} \left\{ S_i^{\dagger}(t) - \hat{S}_i^{\dagger}(t) \right\}^2 dt,
\end{equation}
where $\eta_i$ is the evaluation horizon for the $i$th subject, $S_i^{\dagger}(t)$ is the target survival curve, and $\hat{S}_i^{\dagger}(t)$ is the corresponding estimated curve.
The target $S_i^{\dagger}(t)$ and evaluation horizon $\eta_i$ may vary depending on the objective of the simulation study.
To evaluate efficiency in recovering the unbiased survival function, we set $S_i^{\dagger}(t) = S(t \mid \mathbf{x}_i)$, generate an independent test set of size $n$ without length-biased sampling or right-censoring, and take $\eta_i = \max(\tilde{T}_i)$ \parencite{fu2017survival, fu2017survival_ic}. Alternative choices of $S_i^{\dagger}(t)$ and $\eta_i$ for prediction assessment under realistic LBRC settings are described in Section~3.2.

\begin{figure}[t]
    \centering
    \includegraphics[width=0.75\textwidth]{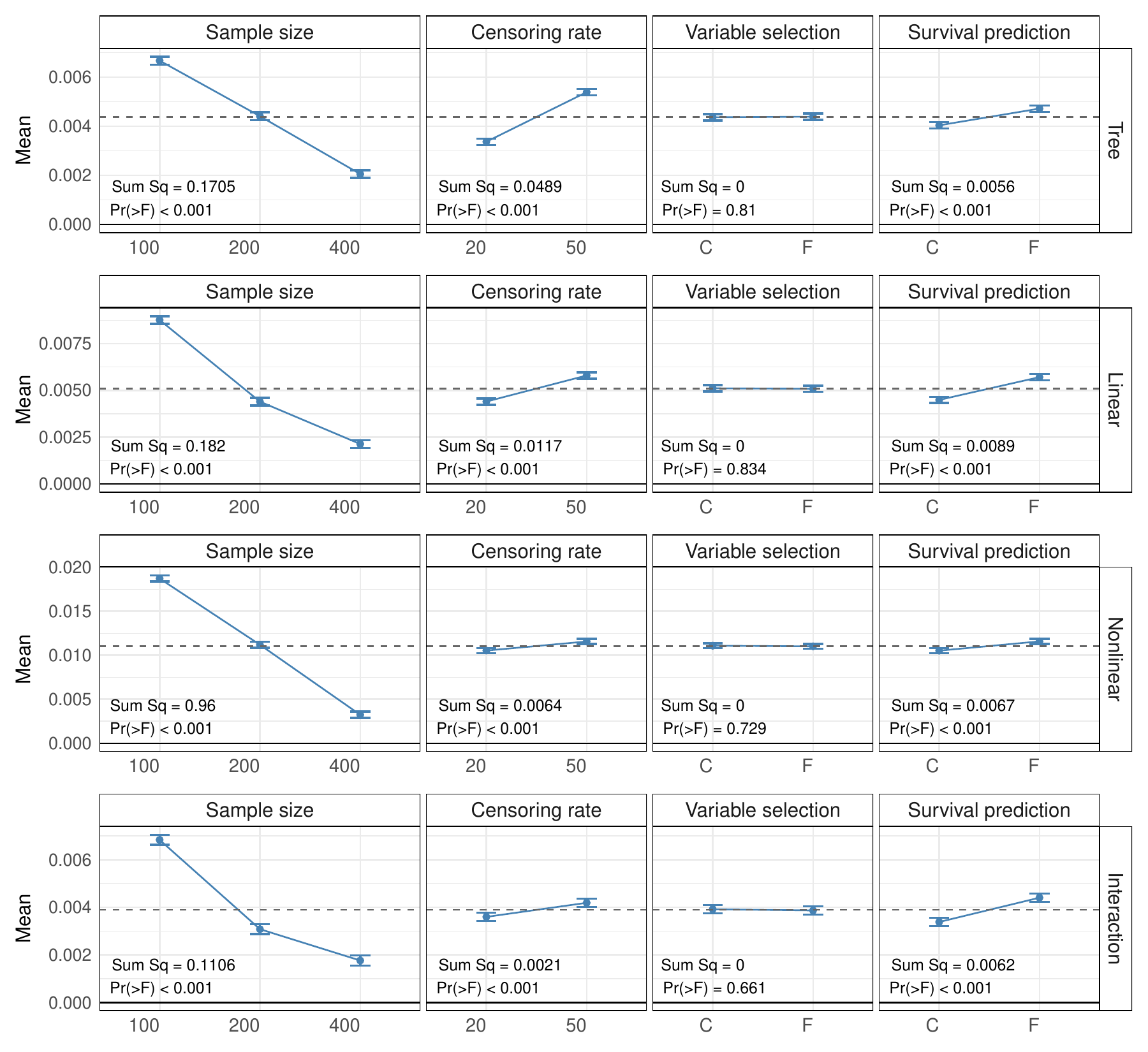}
    \caption{Main effects plots of difference of integrated $L_2$ distance between LTRC-CIT and proposed LBRC-CITs, i.e., $L_2(\text{LTRC-CIT})-L_2(\text{LBRC-CIT})$.
    In each facet, “Sum Sq” denotes the sum of squares attributed to that factor in the ANOVA, and “Pr($>$F)” gives the p-value from the F-test for its main effect.}
    \label{fig3:ANOVA}
\end{figure}

Figure~\ref{fig3:ANOVA} presents main effects plots for the difference of integrated $L_2$-distance between LTRC-CIT and the proposed LBRC-CITs.
The main effect on the failure time distribution is omitted because it is not the focus of this analysis.
The overall positive location of the effects indicates that all four LBRC-CIT variants provide a more accurate estimation of the unbiased survival function than LTRC-CIT. 
Efficiency gains decrease with increasing sample size but become more pronounced under heavier censoring. 
Notably, in the tree-structured scenario, the gain from 20\% to 50\% censoring accounts for a substantial portion of the total main-effect variance. 
In contrast, the effect of censoring is comparatively minor in the linear, nonlinear, and interaction settings. 
One possible explanation is that when the model structure aligns with the data-generating process, LBRC-CIT can more effectively exploit information in the presence of high censoring. 
In misspecified settings, the predictive accuracy is already limited by model fit, reducing the impact of improved censoring adjustment.

Regarding the choice between MCLE and MFLE for variable selection, no meaningful differences in estimation performance are observed. 
This reflects their near-identical tree recovery rates in tree-structured data (Figure~\ref{fig2:Recovery_Rate}) and also holds in the linear and nonlinear scenarios.
In contrast, when comparing unbiased survival function estimators, MFLE consistently achieves a lower integrated $L_2$-distance than MCLE across all data structures.

Having compared all four LBRC-CIT variants across splitting and prediction strategies, we now focus on two representative versions: LBRC-CIT(F,F) (denoted LBRC-CIT-F) and LBRC-CIT(C,C) (denoted LBRC-CIT-C).

\subsection{Properties of the LBRC-CIFs}
We assessed the performance of LBRC-CIF-F and LBRC-CIF-C, which consist of ensembles of LBRC-CIT-F and LBRC-CIT-C trees, respectively. 
All forest models, including LTRC-CIF as a benchmark, are constructed using 100 bootstrap trees. 
The data-generating structures, censoring rates, sample sizes, and number of simulation replicates are identical to those described in Section~\ref{sec:pred_acc}. 

Unlike the study of LBRC-CIT properties, which focused on recovery of the underlying unbiased survival function, our goal here is to evaluate prediction performance under realistic settings, where both length-biased sampling and right-censoring are present. Accordingly, the independent test datasets are generated under the same data-generating mechanism as the training data.
In the integrated \(L_2\)-distance defined in \eqref{eq:L2}, we therefore take \(S_i^\dagger(t)\) as the conditional survival function in \eqref{eq:target_conditional} and set the evaluation horizon to \(\eta_i = Z_i\) to reflect the observable follow-up period in practice \parencite{yao2022ensemble}.
This prediction criterion is used consistently across all competing methods considered in Section~3.2.

\subsubsection{Regulating the construction of trees in forests}
We investigate the sensitivity of the proposed LBRC-CIF methods to the choice of the \texttt{mtry} parameter and evaluate the effectiveness of out-of-bag tuning using the integrated Brier score (IBS) criterion. 
Specifically, we consider the proposed LBRC-CIF-C and LBRC-CIF-F methods, along with the benchmark LTRC-CIF method, within a common out-of-bag tuning framework that targets prediction accuracy for the observed survival outcome. For each CIF, we evaluate prediction performance across \texttt{mtry} values $\{1,2,3,6,12,24,30\}$ and the tuned \texttt{mtry}$,$ using the mean integrated $L_2$-distance relative to the best-performing candidate.

\begin{figure}[t]
    \centering
    \includegraphics[width=\textwidth]{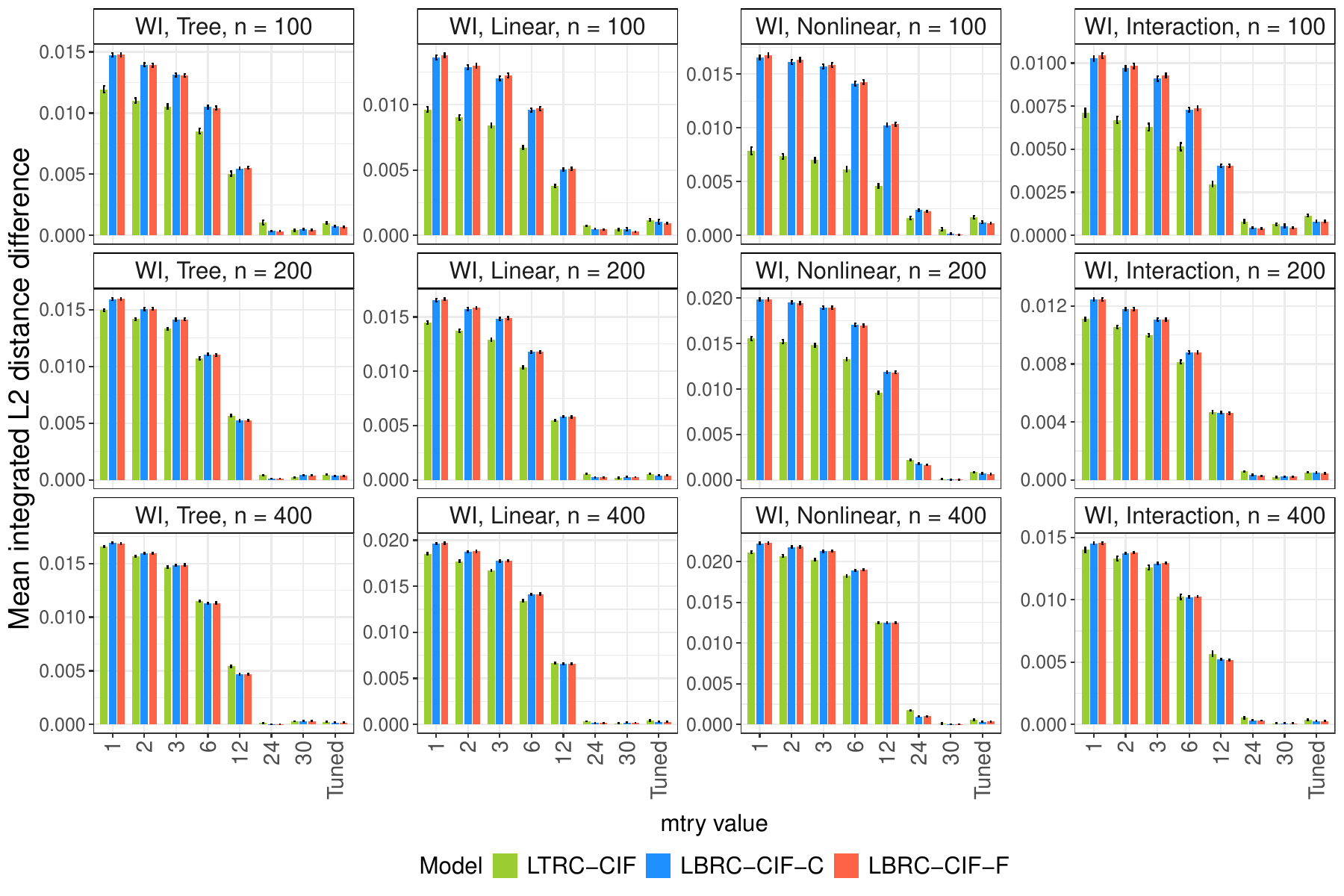}
    \caption{Bar plots of the mean difference from the minimum integrated $L_2$-distance (±1$\cdot$SE) for three CIF variants across \texttt{mtry} candidates $\{1,2,3,6,12,24,30\}$, including the “Tuned” value selected by our IBS-based procedure. Panels show results under Weibull increasing (WI) hazards with 20\% censoring, for the four structures (tree, linear, nonlinear, interaction). $n=100,200,400$.}
    \label{fig4:OOB_tune}
\end{figure}

Figure~\ref{fig4:OOB_tune} illustrates the error bars of these distance differences for the three CIFs under Weibull increasing hazard distributions with 20$\%$ censoring, across different data structures and sample sizes.
Additional results for Weibull decreasing, log-normal, and bathtub-shaped hazard settings are provided in the Supplementary Material. 
In this example, the smallest $L_2$-distance for all CIFs are often attained at \texttt{mtry} values $24$ or $30$, whereas the default value (\texttt{mtry}$=\sqrt{30}\approx6$) does not always perform well.
The small gap between the tuned \texttt{mtry} and the best-performing candidate suggests that our tuning procedure provides a reliable selection strategy.
Notably, for \texttt{mtry} values $1,2,3,6,$ and $12$, the deviation from the minimum is generally larger for LBRC-CIFs than for LTRC-CIF. These findings suggest that the proposed LBRC-CIFs may be more sensitive to the choice of \texttt{mtry}, highlighting the importance of careful tuning.
Based on these findings, we focus on the following analysis on forest methods using the proposed tuned parameter settings.

Similar qualitative patterns were observed when tuning was based on the out-of-bag C-index, indicating that the overall tuning behavior of the proposed LBRC-CIF methods is reasonably robust across different tuning criteria; corresponding results are provided in the Supplementary Material.

\begin{figure}[t]
    \centering
    \includegraphics[width=\textwidth]{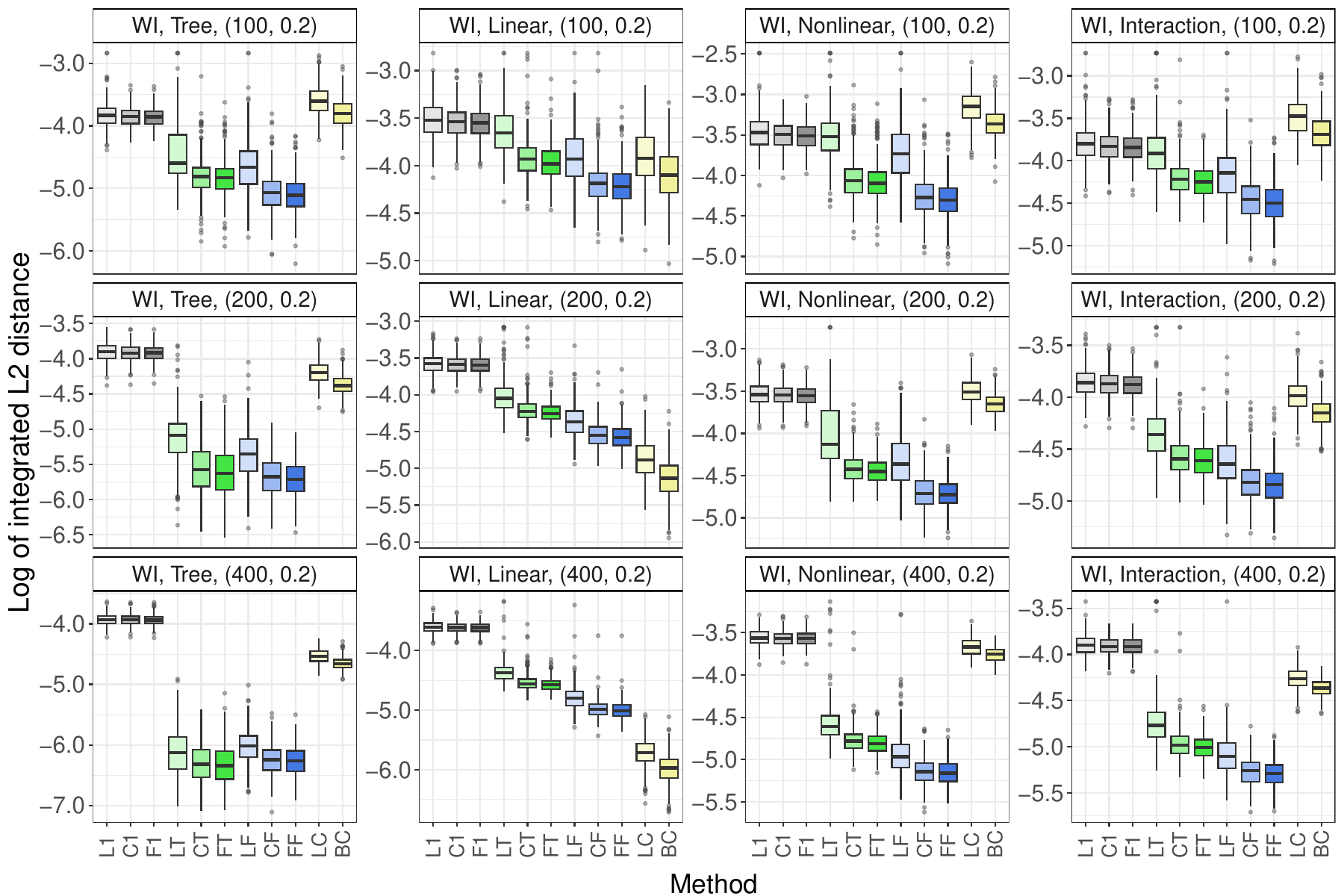}
    \caption{Log-scale integrated $L_2$-distance under Weibull increasing (WI) hazard with 20\% censoring, for the four structures (tree, linear, nonlinear, interaction) and $n=100,200,400$; Panel labels denote $(n,c)$. Grey, green, blue, and yellow boxes denote one-sample estimators, CITs, CIFs, and Cox models, respectively, with lighter to darker shades corresponding to LTRC, MCLE, and MFLE variants within each method class. Extreme outliers beyond the upper whisker are capped.}
    \label{fig5:L2_error}
\end{figure}

\subsubsection{Prediction accuracy across methods}
We evaluate the predictive accuracy of the proposed LBRC forests, LBRC-CIF-F (FF) and LBRC-CIF-C (CF), in comparison to a variety of benchmark methods. 
Competing approaches include the LTRC forest with the same out-of-bag tuning as ours (LF), tree-based methods such as LTRC-CIT (LT), LBRC-CIT-F (FT), and LBRC-CIT-C (CT), and two Cox models: the extended Cox model for left-truncated data (LTRC-COX, LC) and the Cox model for length-biased data using composite likelihood estimation (LBRC-COX, BC; \parencite{huang2012composite}). 
We also include one-sample estimators, which correspond to the stumps (i.e., single-node trees) of the respective tree-based methods: LTRC NPMLE (L1, \parencite{wang1991nonparametric}), MFLE (F1, \parencite{vardi1989multiplicative}), and MCLE (C1, \parencite{he2020nonparametric}).

Figure~\ref{fig5:L2_error} presents side-by-side boxplots of the log-scale integrated $L_2$-distance for all competing methods across tree, linear, nonlinear, and interaction structured setups with Weibull increasing distribution, evaluated at censoring rates of 20\% and sample sizes of $n=100$, 200, and 400. 
Comparable results for other distribution setups are provided in the Supplementary Material. 
At the largest sample size $n=400$, the best performing methods align with the underlying data structure: CITs perform best in tree-structured settings, Cox models lead in linear settings, and CIFs achieve the lowest prediction error in both nonlinear and interaction settings. One-sample estimators perform worst across all scenarios.
Across all methods---CITs, CIFs, and Cox regressions---the LBRC variants consistently outperform their LTRC counterparts. 
Differences between MCLE- and MFLE-based methods are generally modest, although MFLE-based variants occasionally exhibit slightly lower prediction error under heavier censoring or Weibull decreasing hazards, as seen in the supplementary results.

In linear, nonlinear, and interaction settings, LBRC-CIFs generally yield lower prediction error than LBRC-CITs, and likewise, LTRC-CIFs outperform LTRC-CITs, reflecting the expected improvement from ensemble aggregation.
With a smaller sample size \(n=100\), LBRC-CIT methods often outperform the tuned LTRC-CIF across linear, nonlinear, and interaction scenarios. 
This suggests that explicitly accounting for length-biased sampling may sometimes yield greater gains than ensemble aggregation when data are limited.

Overall, the simulation studies showed that incorporating the length-biased sampling structure improves both tree recovery and prediction performance relative to existing LTRC-based methods. The proposed LBRC-CIT methods more accurately recovered the underlying tree structure, while the LBRC-based prediction methods, including the properly tuned forest approaches, generally outperformed their corresponding LTRC counterparts across diverse settings.

We further assessed the robustness of the proposed methods to violations of the stationarity assumption underlying length-biased sampling — that is, to misspecification of the sampling process — considering both homogeneous (a common non-uniform sampling process shared by all subjects) and covariate-dependent misspecification (a sampling process that varies with covariate values). Under mild misspecification, LBRC-CIT and LBRC-CIF retained approximately unbiased variable selection and continued to recover the underlying tree structure more accurately than their LTRC counterparts, consistent with the robustness patterns reported by \textcite{ning2010non} for two-sample testing. Under more severe misspecification, prediction accuracy deteriorated, with the loss concentrated in terminal-node survival estimation rather than split selection, since the splitting score is largely insensitive to the truncation model whereas the node-level estimators depend on it directly. Full simulation details and results are provided in the Supplementary Material.

From a computational perspective, MFLE- and MCLE-based methods showed little difference in runtime for LBRC-CIT, whereas larger differences emerged for the forest methods. In particular, LBRC-CIF-F required repeated EM-based estimation across bootstrap samples and took approximately 10 minutes at $n = 800$, whereas LBRC-CIF-C typically completed in about three minutes.\footnote{All computations were performed on a 64-bit Windows 11 Pro system with an Intel Core i7-13700 processor and 32~GB RAM.} Given the generally modest performance differences between MCLE- and MFLE-based methods at larger sample sizes, LBRC-CIF-C may be preferable to LBRC-CIF-F in large-scale applications.

\section{Real Data Application} \label{sec:application}
\begin{figure}[t]
    \centering
    \includegraphics[width=0.6\textwidth]{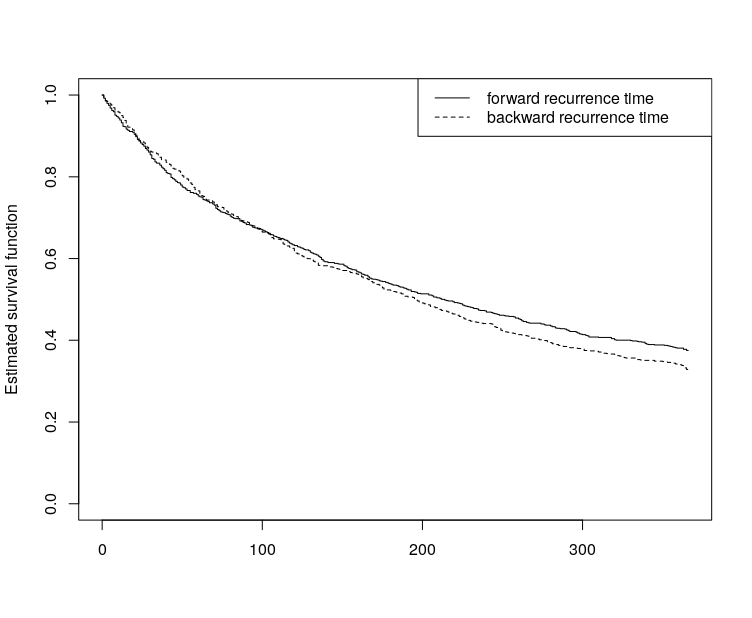}
    \caption{Two Kaplan-Meier curves of backward recurrence time and forward recurrence time among the eligible cohort.}
    \label{fig6:stationary}
\end{figure}
We applied the proposed LBRC-CIT and LBRC-CIF methods to a retrospective cohort study using data from the Cancer Public Library Database (CPLD, \parencite{choi2024data}), a comprehensive cancer registry established in 2022 under the K-CURE project in South Korea. 
The study population consists of lung cancer patients with histologically confirmed diagnoses between January 1, 2012 and December, 31, 2019. 
To ensure at least one full year of healthcare utilization data prior to diagnosis for accurate assessment of comorbidities, patients diagnosed in 2012 are excluded. 
Our focus is on the survival of patients with clinically serious lung cancer who did not initiate treatment soon after diagnosis.
Accordingly, the eligible cohort met the following criteria:
\begin{enumerate}
    \item Age ≥ 65 years at diagnosis;
    \item Distant stage, defined by the SEER summary staging system;
    \item No lung cancer treatment (surgery, radiotherapy, or chemotherapy) within four months of diagnosis.
\end{enumerate}
We collected baseline covariates, including sex, body mass index, moderate or vigorous physical activity, five blood test measures, and smoking status. Additionally, three respiratory comorbidities, fifteen other medical comorbidities, and six histologic subtypes were included, resulting in a total of 34 covariates. 
Baseline covariates were defined using only information available at the time of lung cancer diagnosis to ensure that they reflected patients’ clinical status at cohort entry. The baseline time for survival analysis was defined as the date of lung cancer diagnosis.
Detailed information on CPLD and variables used for eligibility criteria and covariate definitions is provided in the Supplemental Material.

\begin{figure}[h!]
    \centering
    \includegraphics[width=0.53\textwidth]{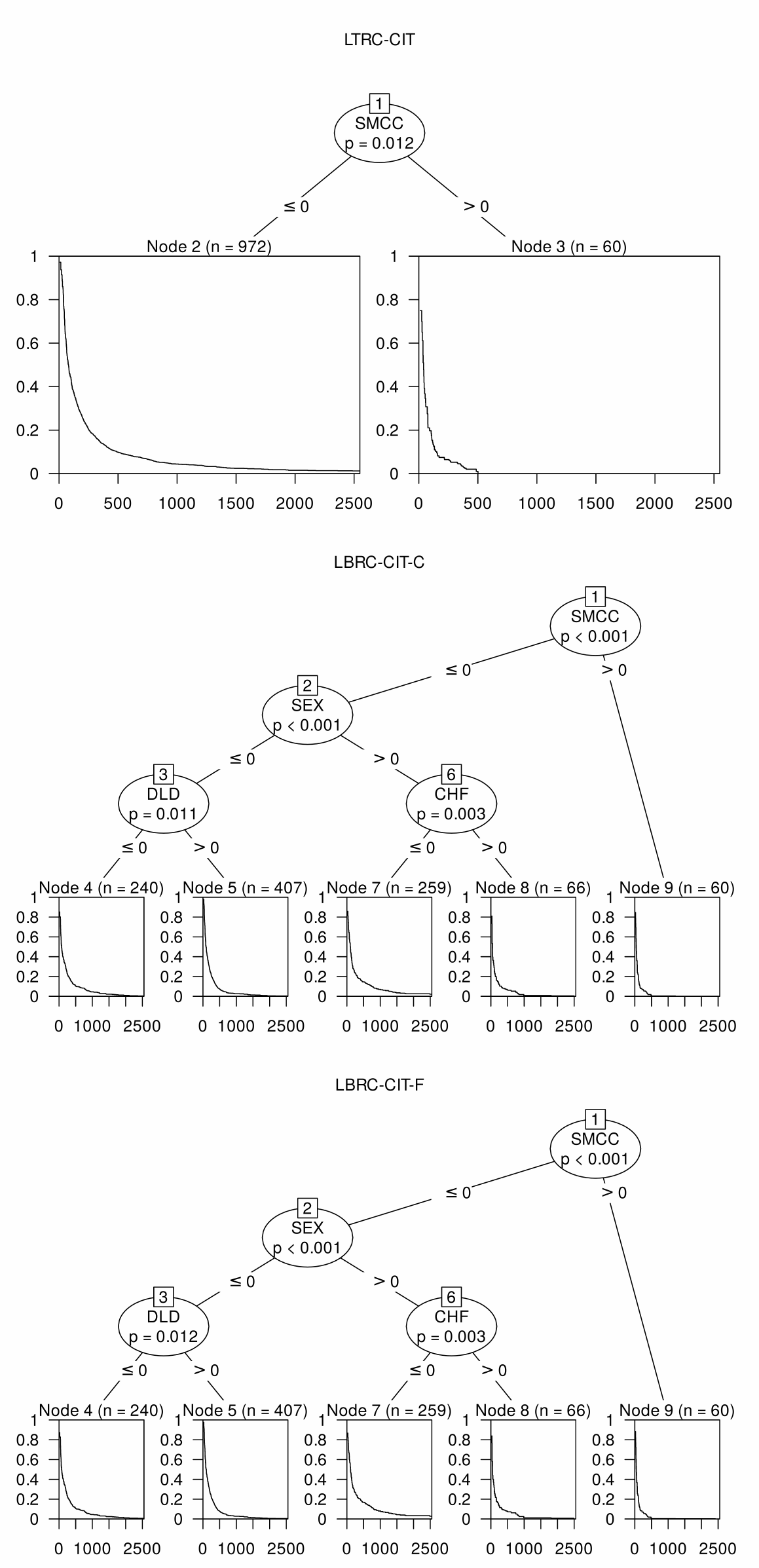}
    \caption{ CIT results for LTRC-CIT (top), LBRC-CIT-C (middle), and LBRC-CIT-F (bottom). SMCC, CHF, DLD denote small cell carcinoma, congestive heart failure, and dyslipidemia; for binary covariates, 1 indicates presence (female for SEX) and 0 absence (male).}
    \label{fig7:tree_plots}
\end{figure}

We defined January 1, 2020, as the cohort entry date and considered a 1-year follow-up period thereafter.
From 11,060 newly diagnosed patients between January 1, 2013, and December 31, 2019, we identified 1,032 who were alive at the entry date.
Using claims records, we then ascertained death dates or censoring through December, 31, 2020.
By the end of this period, 645 patients had died, and the remainder were right-censored.

To assess the stationarity assumption of the truncation time, we compared the distribution of time from diagnosis to enrollment with that of time from enrollment to event using the method of \textcite{addona2006formal} (R package: \textcite{lee2020coxphlb}).
The resulting p-value was 0.491, indicating no significant deviation from the stationarity assumption. 
Additionally, we evaluated the assumption graphically by comparing the Kaplan-Meier estimates for the backward and truncation times as suggested by \textcite{asgharian2006checking}. 
As shown in Figure~\ref{fig6:stationary}, the two survival curves were nearly indistinguishable, further supporting the stationarity of the truncation time.

\begin{figure}[t]
    \centering
    \includegraphics[width=0.9\textwidth]{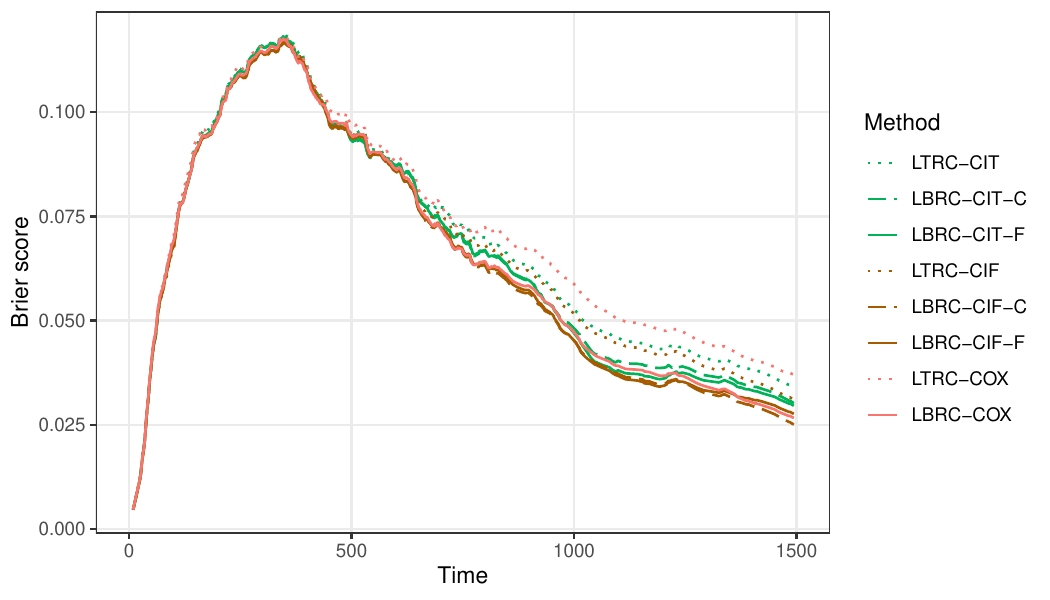}
    \caption{Mean Brier score over follow-up time($10\times10$ repeated cross-validation), shown up to the last time point at which at least 10\% of subjects remain at risk. LTRC-COX and LBRC-COX denote the extended and length-biased Cox models, respectively.}
\label{fig8:brier_score}
\end{figure}

We fit the proposed LBRC-CIT and LBRC-CIF methods along with the LTRC-CIT, LTRC-CIF, LTRC-COX, and LBRC-COX methods. To evaluate predictive performance, we computed \(10\times10\) repeated cross-validation errors (i.e., 10-fold cross-validation repeated 10 times, averaging the Brier score over each held-out fold) at each follow-up day. The prediction errors and their integrated summaries were evaluated up to the time point at which approximately 10\% of subjects remained at risk. The corresponding integrated summaries are reported in Table~\ref{tab:ibs}.

\begin{table}[H]
    \centering
    \begin{tabular}{lcccccccc}
    \toprule
     & \multicolumn{3}{c}{CIT} & \multicolumn{3}{c}{CIF} & \multicolumn{2}{c}{COX} \\
    \cmidrule(lr){2-4} \cmidrule(lr){5-7} \cmidrule(lr){8-9}
    IBS & LTRC & LBRC-C & LBRC-F & LTRC & LBRC-C & LBRC-F & LTRC & LBRC \\
    \midrule
    Mean & 0.0687 & 0.0661 & 0.0656 & 0.0672 & 0.0640 & 0.0641 & 0.0709 & 0.0648 \\
    (SD) & (0.0066)& (0.0064)& (0.0054)& (0.0062)& (0.0054)& (0.0053)& (0.0076)& (0.0055) \\
    \bottomrule
    \end{tabular}
    \caption{Mean integrated Brier score with standard deviations in parentheses.}
    \label{tab:ibs}
\end{table}

Figure~\ref{fig7:tree_plots} presents the fitted trees for LTRC-CIT, LBRC-CIT-C, and LBRC-CIT-F. All three trees initially split on histologic subtype (small-cell carcinoma), with LTRC-CIT exhibiting the weakest statistical evidence. Reflecting its highest IBS, LTRC-CIT appears less sensitive to secondary prognostic factors. Both LBRC variants then split on sex, with congestive heart failure selected on the right child node and dyslipidemia on the left. The LBRC-CIT variants exhibit lower IBS values than LTRC-CIT, with LBRC-CIT-F showing slightly improved performance and lower variability than LBRC-CIT-C.

Figure~\ref{fig8:brier_score} displays the mean Brier score trajectories over follow-up time. 
Across follow-up time, the LBRC-based methods maintain lower prediction error than their corresponding LTRC-based counterparts, with the differences becoming more apparent during later follow-up periods.
These trends are consistent with the integrated Brier score summaries in Table~\ref{tab:ibs}. 
Among the tree-based approaches, LBRC-CIF-C and LBRC-CIF-F achieve the lowest IBS values, followed by the LBRC-CIT variants. 
The LBRC-COX model also demonstrates competitive prediction performance, whereas the LTRC-based methods generally exhibit higher IBS values. 
The nearly identical performance of the two LBRC-CIF variants suggests that the computationally simpler MCLE approach may achieve prediction accuracy comparable to that of the MFLE approach in this application.

\section{Discussion} \label{sec:discussion}
In this paper, we propose novel tree-based methods for analyzing length-biased right-censored data. 
Our approach addresses the inefficiency of existing conditional inference trees developed for left-truncated data by incorporating score functions based on the full likelihood and nonparametric survival estimators tailored to length-biased settings. 
We introduced two variants of the LBRC-CIT: a fully efficient version and a closed-form approximation, offering a balance between statistical performance and computational scalability. 
Simulation studies demonstrated that the proposed LBRC-CIT and LBRC-CIF generally achieved more reliable tree recovery and improved prediction performance than their LTRC counterparts across a wide range of settings, particularly when the forest tuning parameters were appropriately selected.
Sensitivity analyses further suggested that the proposed methods remained robust to moderate departures from the stationarity assumption, maintaining competitive performance in tree recovery and prediction.
In the real data application, the LBRC-based methods also generally demonstrated improved prediction performance relative to their corresponding LTRC-based counterparts.

Although the integrated Brier score and C-index are widely used for tuning and evaluating survival prediction models, recent work has advocated predictive likelihood as a principled alternative because survival models naturally produce an entire predictive survival distribution rather than a single point estimate \cite{lu2025model}. Direct use of predictive likelihood is challenging for semiparametric and nonparametric survival models because their step-function survival estimators assign zero likelihood to event times not observed in the training data; however, \textcite{lu2025model} addressed this issue using nearest-neighbor smoothing, thereby making predictive likelihood applicable to more general semiparametric and nonparametric survival models. Since both LBRC-CIF and LTRC-CIF rely on nonparametric step-function survival estimators, smoothing-based predictive likelihood criteria may provide a useful extension for tuning and evaluating LBRC and LTRC forest models.

At the same time, predictive likelihood construction under length-biased sampling raises important questions regarding the target prediction objective. For general left-truncated data, one natural approach is to construct predictive likelihoods conditional on the truncation time, corresponding to predicting the observable survival distribution after study entry. Under length-biased sampling, one may also incorporate the truncation mechanism itself into a full predictive-likelihood formulation. Alternatively, truncation-independent predictive likelihoods that target the underlying unbiased survival distribution in the underlying disease population may be considered \cite{hartman2023concordance}. These formulations correspond to distinct scientific objectives and should therefore be studied under target-specific evaluation designs. For observed-data prediction after study entry, conditional and full predictive likelihood criteria could be compared in terms of tuning stability and efficiency gains under length-biased sampling. In contrast, truncation-independent predictive likelihood is more naturally aligned with recovery of the underlying unbiased survival distribution in the underlying disease population and could be evaluated through inverse-truncation-weighted validation schemes or simulation-based target assessments.

Beyond the choice of tuning criteria, another important direction is to extend the underlying tree and forest framework beyond proportional hazards settings, since conditional inference trees and forests rely on log-rank-based splitting criteria.
For non-proportional hazards, the transformation forest framework of \textcite{hothorn2021predictive} offers a promising alternative and has already been adapted to left-truncated data by \textcite{yao2022ensemble}. Extending transformation forests to length-biased right-censored data, therefore, represents a natural and promising direction for future methodological development.

\subsubsection*{Acknowledgments} This work was supported by the National Cancer Center of Korea (Grant Nos. NCC-2410690-2 and NCC-2511590-1).

\subsubsection*{Ethics approval and consent to participate} This study followed the ethical standards of the Declaration of Helsinki, with approval (IRB no. NCC-2023-0272) from the Institutional Review Board of the National Cancer Center of Korea, which waived the need for written informed consent.

\subsubsection*{Conflicts of Interest} The authors declare no conflicts of interest.

\subsubsection*{Data Availability Statement} The R code and intermediate simulation results required to reproduce the analyses are provided in the accompanying Supplement. The R package implementing the proposed LBRC-CIT and LBRC-CIF methods is available at \url{https://github.com/jinwu99/LBRCtreeforests}. The real data obtained from the Cancer Clinical Library Database are not publicly available due to privacy regulations.

\printbibliography

\end{refsection}

\clearpage

\begin{refsection}

\setcounter{page}{1}
\renewcommand{\thepage}{S\arabic{page}}

\setcounter{table}{0}
\renewcommand{\thetable}{S\arabic{table}}
\setcounter{figure}{0}
\renewcommand{\thefigure}{S\arabic{figure}}
\setcounter{section}{0}
\renewcommand\thesection{\Alph{section}}

\begin{center}
{\large\bf SUPPLEMENTAL MATERIALS}
\end{center}

\section{Derivation of Vardi's discretized hazard function}
To derive $\lambda_j$, we use the expected complete-data log-likelihood of \textcite{qin2011maximum}, which leads to the same NPMLE as \textcite{vardi1989multiplicative} but provides a more convenient derivation.
Let $p_j = dF(t_j)$ denote the unbiased failure mass at $t_j$. 
With the cumulative hazard $\Lambda(u)=\sum_{u\geq t_j} \lambda_j$, the $p_j$ can be expressed as $\lambda_j \exp \left( - \sum_{k=1}^j \lambda_k\right)$, so the expected complete-data log-likelihood (equation (4) in \textcite{qin2011maximum}) becomes
\begin{equation*}
    l_E(\lambda) = \sum_{j=1}^h w_j \log \lambda_j - \sum_{j=1}^h \sum_{k=j}^h \lambda_j w_k. 
\end{equation*}
where $w_j$ are the E–step weights.
Maximizing this expression with respect to $\lambda_j$ yields
\begin{equation*}
    \lambda_j = \frac{w_j}{\sum_{k=j}^h w_k}. 
\end{equation*}
At convergence of the EM algorithm, each weight $w_j$ is proportional to $\hat p_j$, and using the length-bias relation $\hat p_j = t_j^{-1} \hat q_j / \left( \sum_{k=1}^h t_k^{-1} \hat q_k \right)$ gives
\begin{equation*}
    \hat \lambda_j = \frac{t_j^{-1} \hat q_j}{\sum_{k=j}^h t_k^{-1} \hat q_k},
\end{equation*}
which is the discretized hazard representation of Vardi’s NPMLE.

\clearpage
\section{Test of unbiasedness of variable selection}
The unbiased failure time $\tT$ is randomly generated with the following distributions:
\begin{itemize}
    \item Weibull with increasing hazard, shape parameter $\alpha=2$ and scale parameter $\beta=3$
    \item Weibull with decreasing hazard, shape parameter $\alpha=0.9$ and scale parameter $\beta=2$
    \item Lognormal with mean $\mu=1.4$ and standard deviation $\sigma=0.4$
\end{itemize}
The observed response for each observation is a triplet $(Z,A,\delta)$, where $Z=\min(T,C)$. There are six independent covariates $X_1,...,X_6$, where $X_1,X_2$ randomly take values from the set $\{1,2,3,4,5,6\}$, are $U[0,1]$, $X_3,X_4$ are binary$\{0,1\}$, and $X_5,X_6$ are $U[0,1]$.
Since the response triplet $(Z,A,\delta)$ is generated independently from the covariates $X_1,...,X_6$, there does not exist any true association between the survival outcomes and covariates, so the tree algorithms should not split on any of the covariates; unbiasedness would imply that if they are forced to split with equal probabilities for all six. There are $10,000$ simulation replicates in each setting with sample size $n=200$. 

\begin{figure}[!h]
    \centering
    \includegraphics[width=0.75\textwidth]{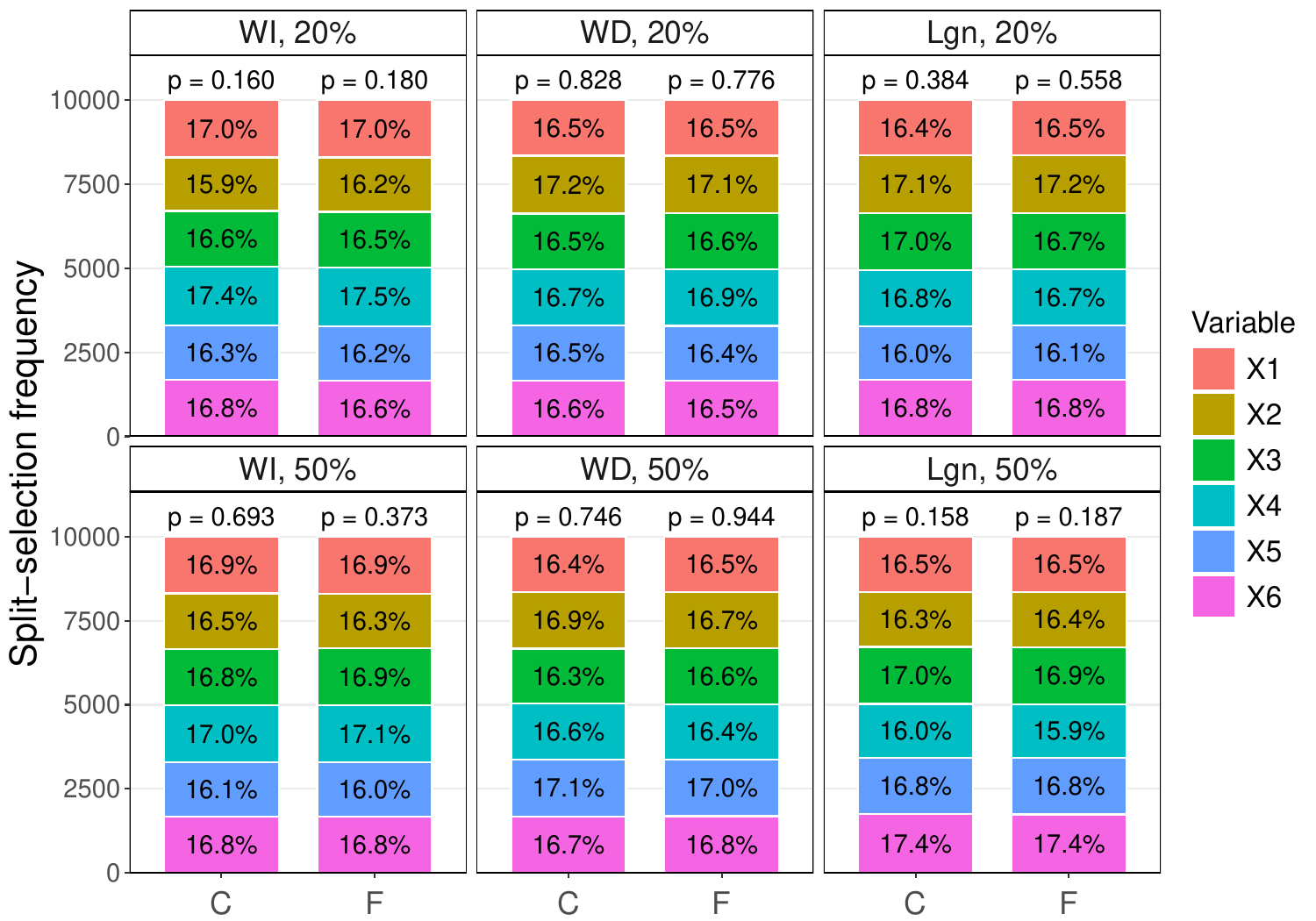}
    \caption{Root split-selection frequencies under null settings. C and F denote LBRC-CIT(C,\(\cdot\)) and LBRC-CIT(F,\(\cdot\)), respectively. Percentages within bars are based on (10{,}000) simulation replicates.}
    \label{figS:Unbias}
\end{figure}

Figure~\ref{figS:Unbias} displays the split-selection frequencies of the root split variable across simulation settings, together with the corresponding Pearson chi-squared test \(p\)-values for equality of split-selection probabilities among the six covariates. Across all settings, both LBRC-CIT(C,\(\cdot\)) and LBRC-CIT(F,\(\cdot\)) exhibit nearly uniform split-selection frequencies and non-significant \(p\)-values, indicating that the proposed methods are approximately unbiased in variable selection.

\clearpage

\section{Regulating the construction of trees in forests}

\subsection{Out-of-bag tuning criteria}

In the main manuscript, Section 2.4 introduces out-of-bag tuning procedures for selecting the \texttt{mtry} parameter in CIF models using the out-of-bag integrated Brier score (IBS) and C-index criteria, while Section 3.2.1 presents the corresponding results under Weibull increasing hazard distributions. Additional tuning results based on the out-of-bag C-index are provided in this supplement to assess robustness across tuning criteria.

We briefly summarize the C-index-based tuning procedure. Let \((Z_i,\delta_i)\) denote the observed survival outcome for subject \(i\), and let \(\hat S_{\x_i}(t)\) denote the predicted conditional survival function defined in equation (17) of the main manuscript. Following the prediction framework of \textcite{yao2022ensemble}, the out-of-bag C-index is computed from comparable subject pairs and used to select the \texttt{mtry} value maximizing concordance between predicted and observed outcomes. Specifically, we use the cumulative risk score
\[
\hat r_i = \int_0^{\max(Z_i)} \left\{ 1-\hat S_{\x_i}(t) \right\}\,dt,
\]
where larger values correspond to higher predicted risk. Concordance is then evaluated over comparable out-of-bag subject pairs under the left-truncated right-censored setting; specifically, a pair \((i,j)\) is considered comparable if subject \(i\) experiences an observed event before subject \(j\) while subject \(j\) remains under observation at that time. The \texttt{mtry} value maximizing the resulting out-of-bag C-index is selected for each forest.

\subsection{Additional tuning results}

\subsubsection{Results based on IBS}

\begin{figure}[!h]
    \centering
    \includegraphics[width=\textwidth]{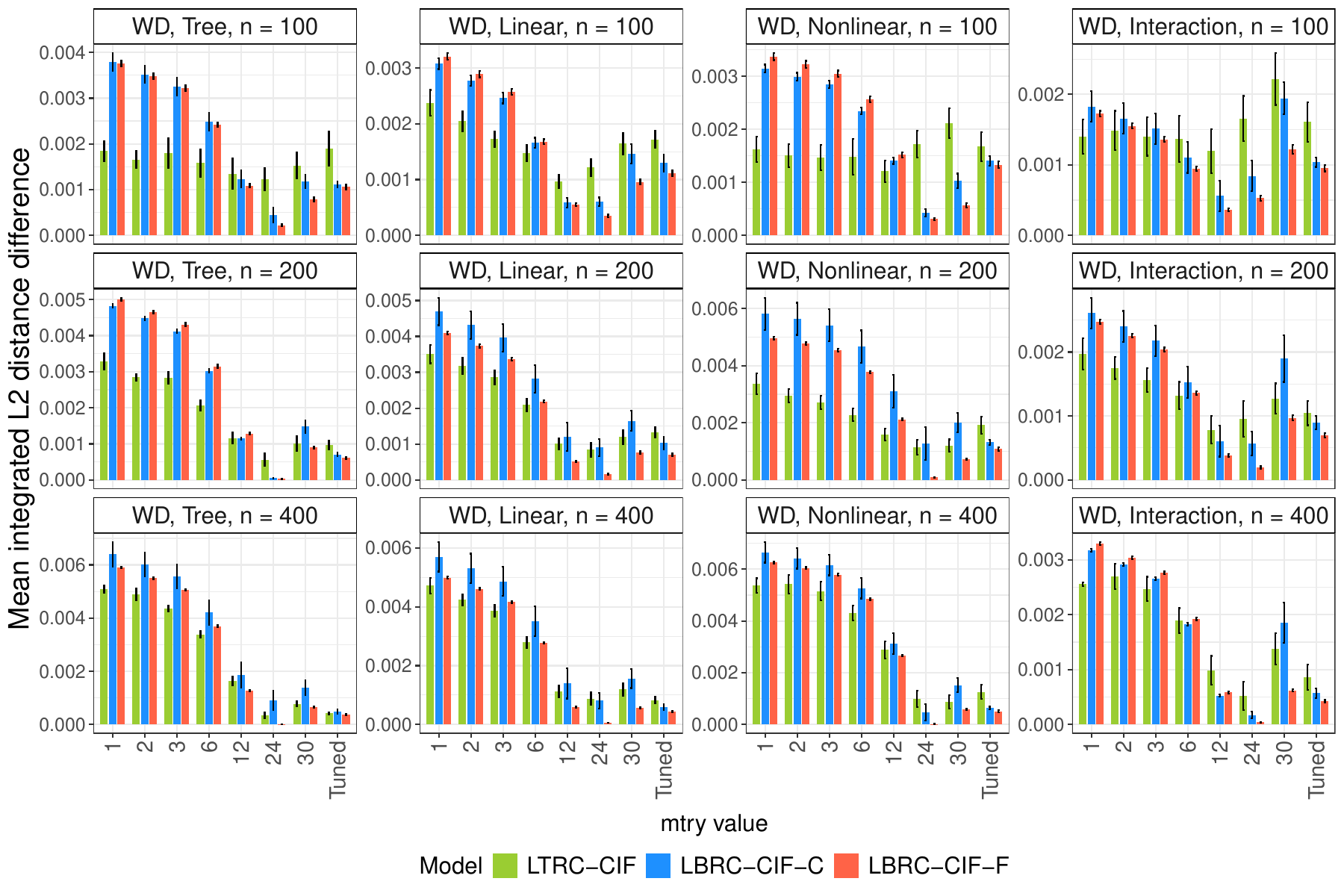}
    \caption{Bar plots of the mean difference from the minimum integrated $L_2$-distance (±1$\cdot$SE) for three CIF variants across \texttt{mtry} candidates $\{1,2,3,6,12,24,30\}$, including the tuned value selected by the IBS-based out-of-bag procedure. Panels show results under Weibull decreasing hazards with 20\% censoring across tree, linear, nonlinear, and interaction structures and sample sizes $n=100,200,400$.}
    \label{figS:regulate_trees_WD}
\end{figure}

\begin{figure}[!h]
    \centering
    \includegraphics[width=0.9\textwidth]{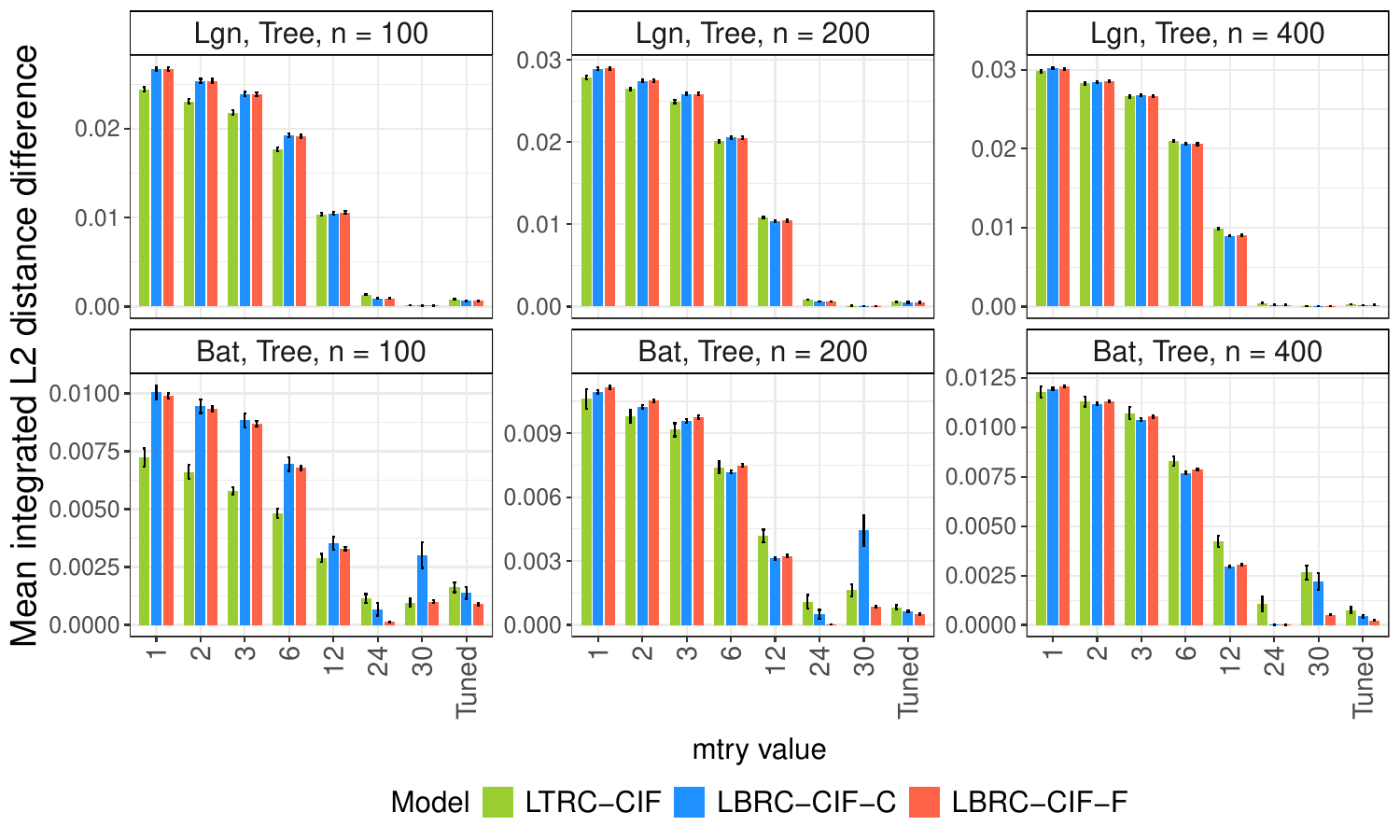}
    \caption{Bar plots of the mean difference from the minimum integrated $L_2$-distance (±1$\cdot$SE) for three CIF variants across \texttt{mtry} candidates $\{1,2,3,6,12,24,30\}$, including the tuned value selected by the IBS-based out-of-bag procedure. Panels show results under log-normal (Lgn) and bathtub-shaped (Bat) hazards with 20\% censoring under tree-structured settings and sample sizes $n=100,200,400$.}
    \label{figS:regulate_trees_LgnBat}
\end{figure}

Figures~\ref{figS:regulate_trees_WD} and~\ref{figS:regulate_trees_LgnBat} present additional IBS-based tuning results under Weibull decreasing, log-normal, and bathtub-shaped hazard settings.
Overall, the additional IBS-based results show qualitative patterns similar to those reported in the main manuscript. In most scenarios, larger \texttt{mtry} values tended to yield lower prediction error, while the tuned \texttt{mtry} values generally achieved performance close to the best-performing candidate. This agreement became more pronounced as the sample size increased, suggesting that the out-of-bag tuning procedure becomes increasingly stable in larger samples. Although somewhat greater variability was observed under Weibull decreasing hazard settings, particularly for extreme \texttt{mtry} values in smaller samples, the overall tuning behavior remained broadly consistent across distributions.

\clearpage

\subsubsection{Results based on the C-index}

\begin{figure}[!h]
    \centering
    \includegraphics[width=\textwidth]{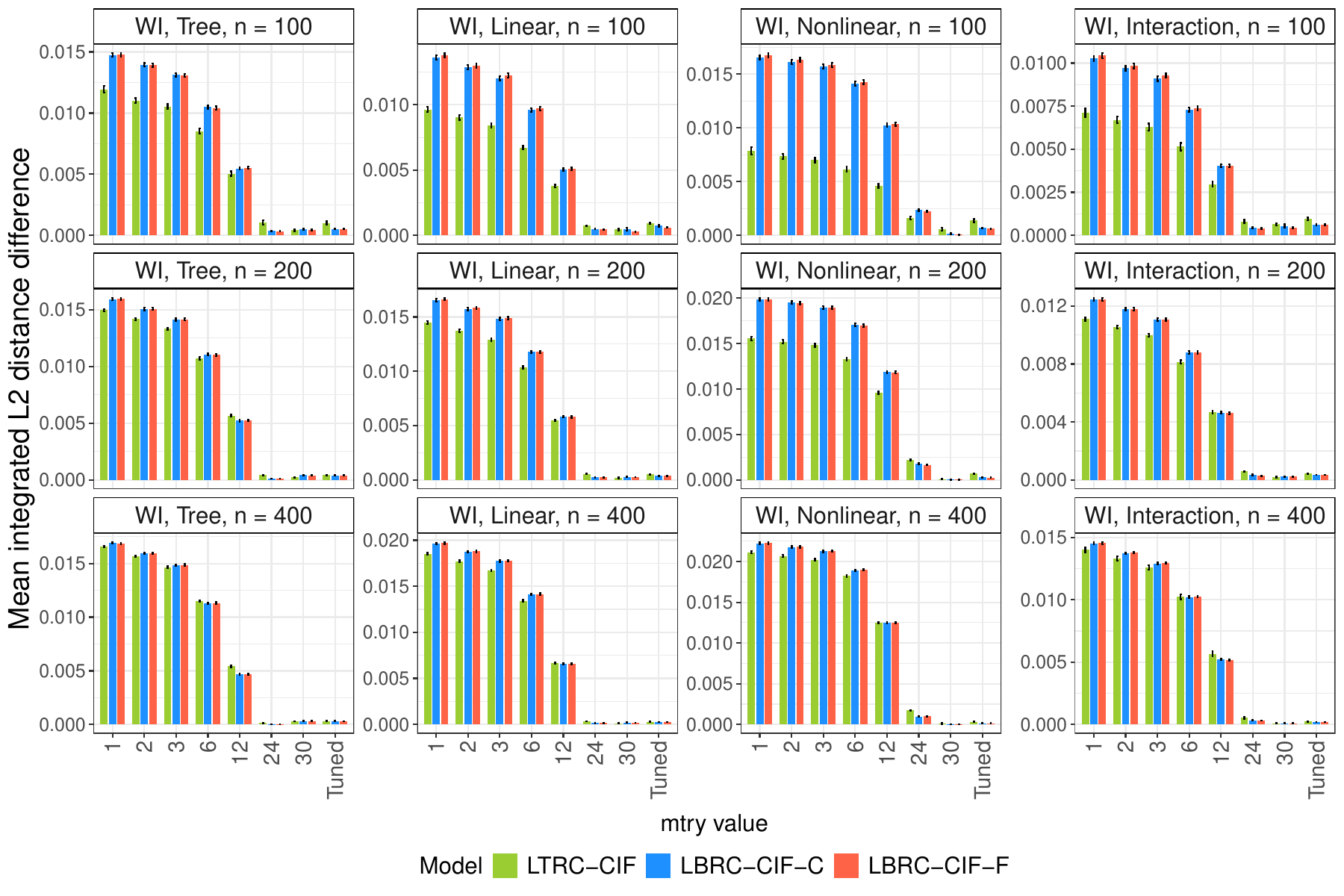}
    \caption{Bar plots of the mean difference from the minimum integrated $L_2$-distance (±1$\cdot$SE) for three CIF variants across \texttt{mtry} candidates $\{1,2,3,6,12,24,30\}$, including the tuned value selected by the C-index-based out-of-bag procedure. Panels show results under Weibull increasing hazards with 20\% censoring across tree, linear, nonlinear, and interaction structures and sample sizes $n=100,200,400$.}
    \label{figS:regulate_trees_WI_C}
\end{figure}

\begin{figure}[!h]
    \centering
    \includegraphics[width=\textwidth]{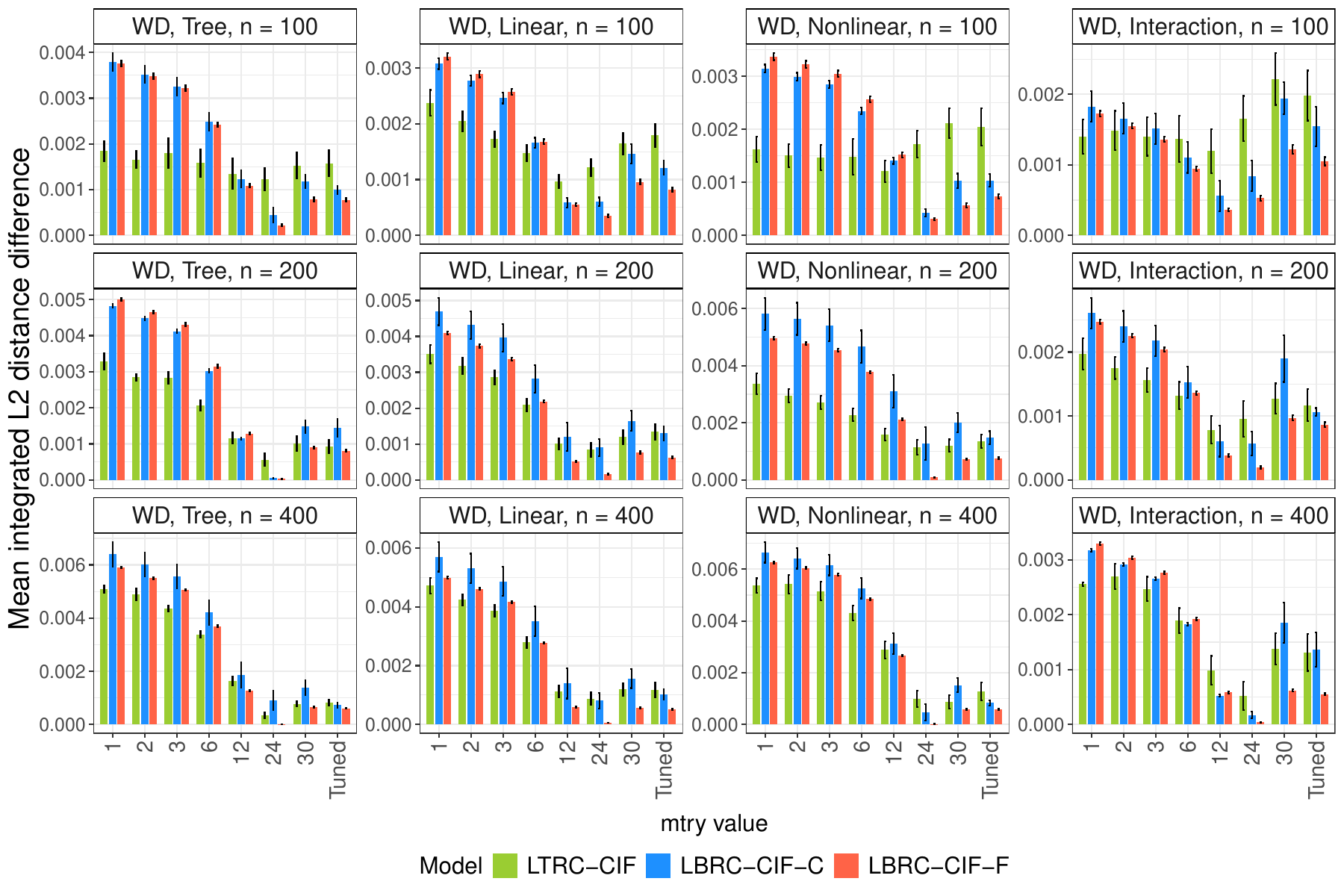}
    \caption{Bar plots of the mean difference from the minimum integrated $L_2$-distance (±1$\cdot$SE) for three CIF variants across \texttt{mtry} candidates $\{1,2,3,6,12,24,30\}$, including the tuned value selected by the C-index-based out-of-bag procedure. Panels show results under Weibull decreasing hazards with 20\% censoring across tree, linear, nonlinear, and interaction structures and sample sizes $n=100,200,400$.}
    \label{figS:regulate_trees_WD_C}
\end{figure}

\begin{figure}[!h]
    \centering
    \includegraphics[width=0.9\textwidth]{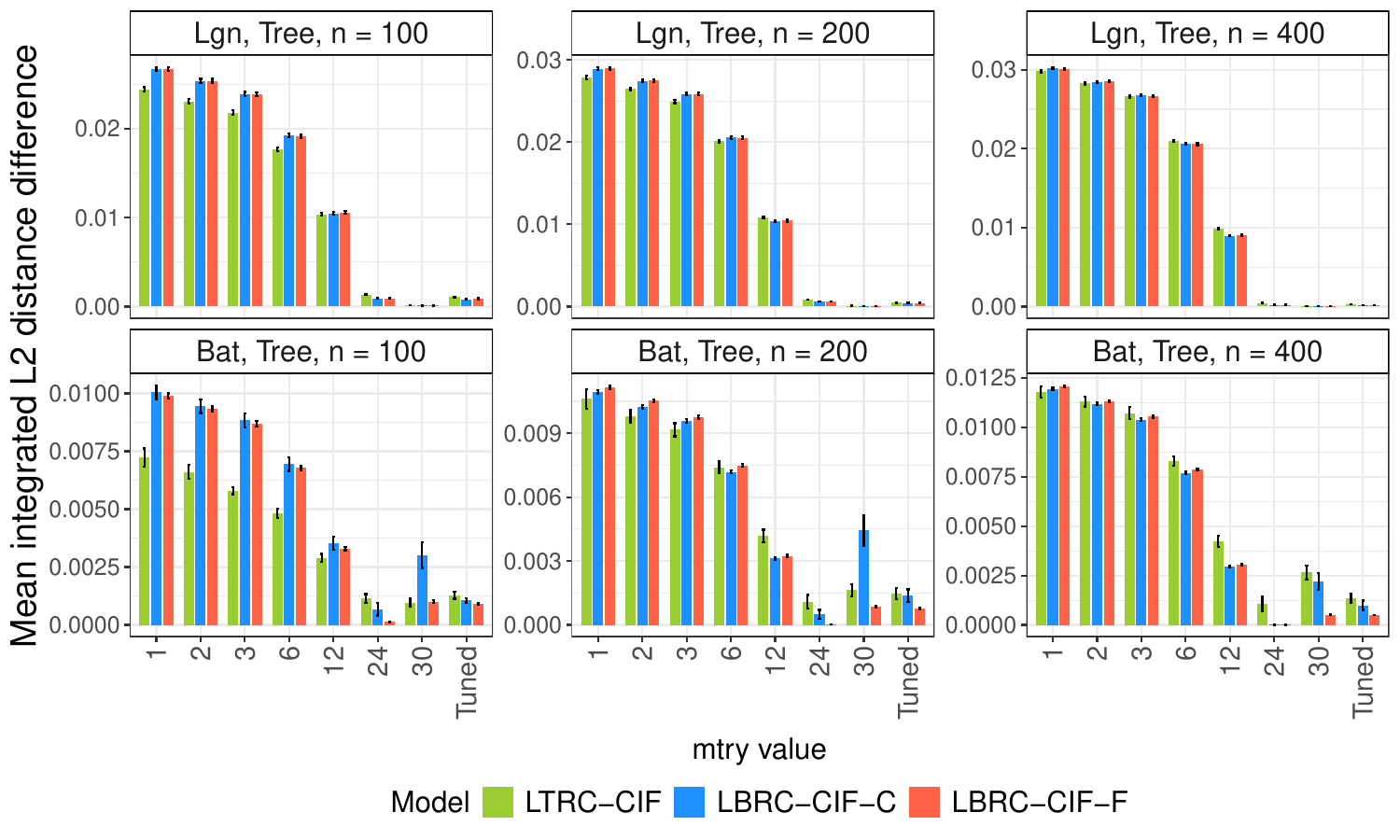}
    \caption{Bar plots of the mean difference from the minimum integrated $L_2$-distance (±1$\cdot$SE) for three CIF variants across \texttt{mtry} candidates $\{1,2,3,6,12,24,30\}$, including the tuned value selected by the C-index-based out-of-bag procedure. Panels show results under log-normal (Lgn) and bathtub-shaped (Bat) hazards with 20\% censoring under tree-structured settings and sample sizes $n=100,200,400$.}
    \label{figS:regulate_trees_LgnBat_C}
\end{figure}

Figures~\ref{figS:regulate_trees_WI_C}--\ref{figS:regulate_trees_LgnBat_C} present corresponding tuning results when the out-of-bag C-index was used as the tuning criterion.
Overall, the C-index-based tuning results were qualitatively similar to those obtained using IBS. Across most settings, the tuned \texttt{mtry} values achieved prediction performance close to the best-performing candidate values, and the tuning stability improved as the sample size increased. Similar to the IBS-based results, lower \texttt{mtry} values generally produced noticeably larger deviations from the minimum prediction error, particularly for the proposed LBRC-CIF methods. Although some additional variability appeared under Weibull decreasing hazards, the overall tuning behavior remained reasonably consistent across tuning criteria.

\clearpage

\section{Prediction accuracy across methods}

\begin{figure}[!h]
    \centering
    \includegraphics[width=\textwidth]{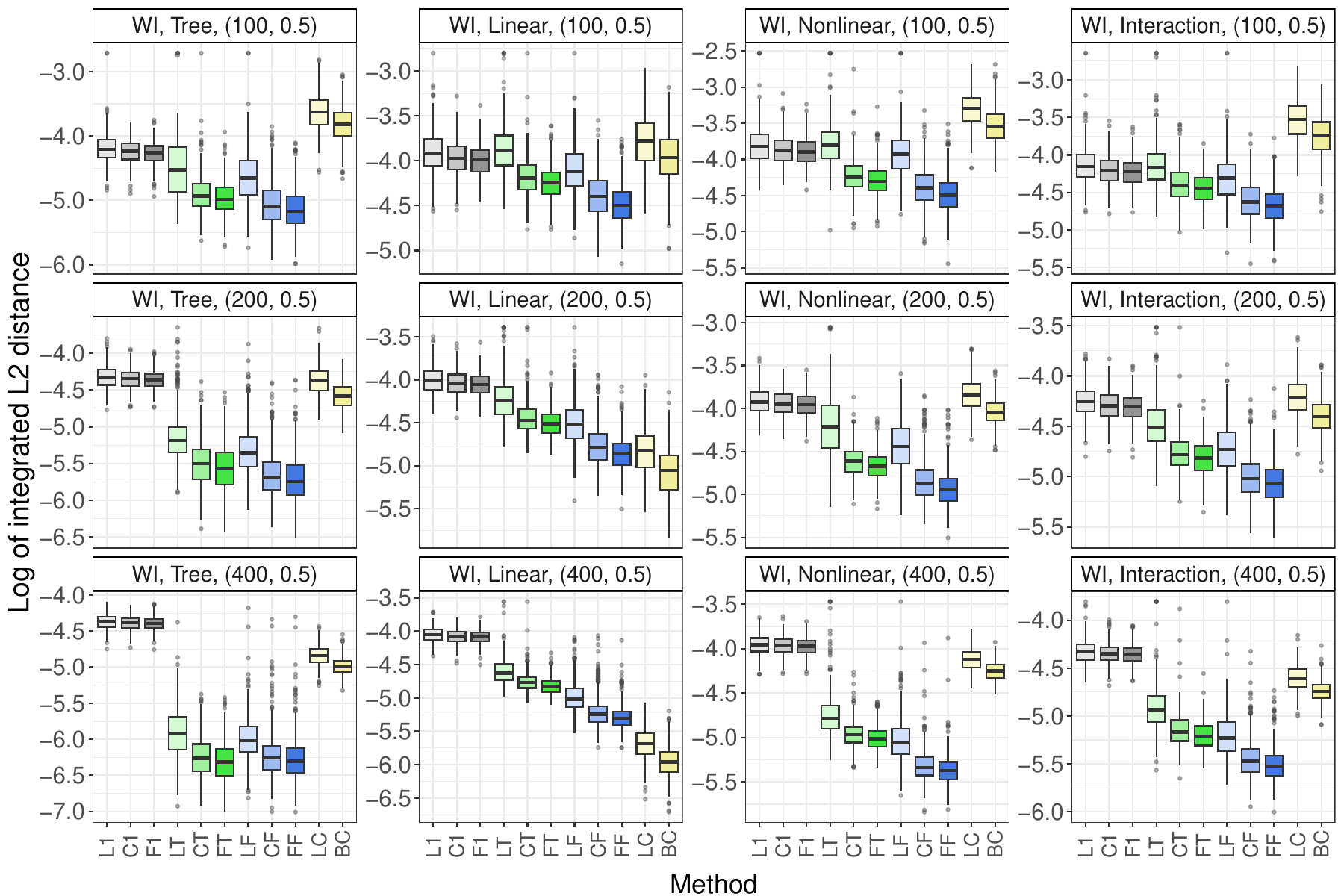}
    \caption{Log-scale integrated $L_2$-distance under Weibull increasing (WI) hazard with 50\% censoring, for the four structures (tree, linear, nonlinear, interaction) and $n=100,200,400$; Panel labels denote $(n,c)$. Grey, green, blue, and yellow boxes denote one-sample estimators, CITs, CIFs, and Cox models, respectively, with lighter to darker shades corresponding to LTRC, MCLE, and MFLE variants within each method class. Extreme outliers beyond the upper whisker are capped.}
    \label{figS:prediction_WI_50}
\end{figure}

\begin{figure}[!h]
    \centering
    \includegraphics[width=\textwidth]{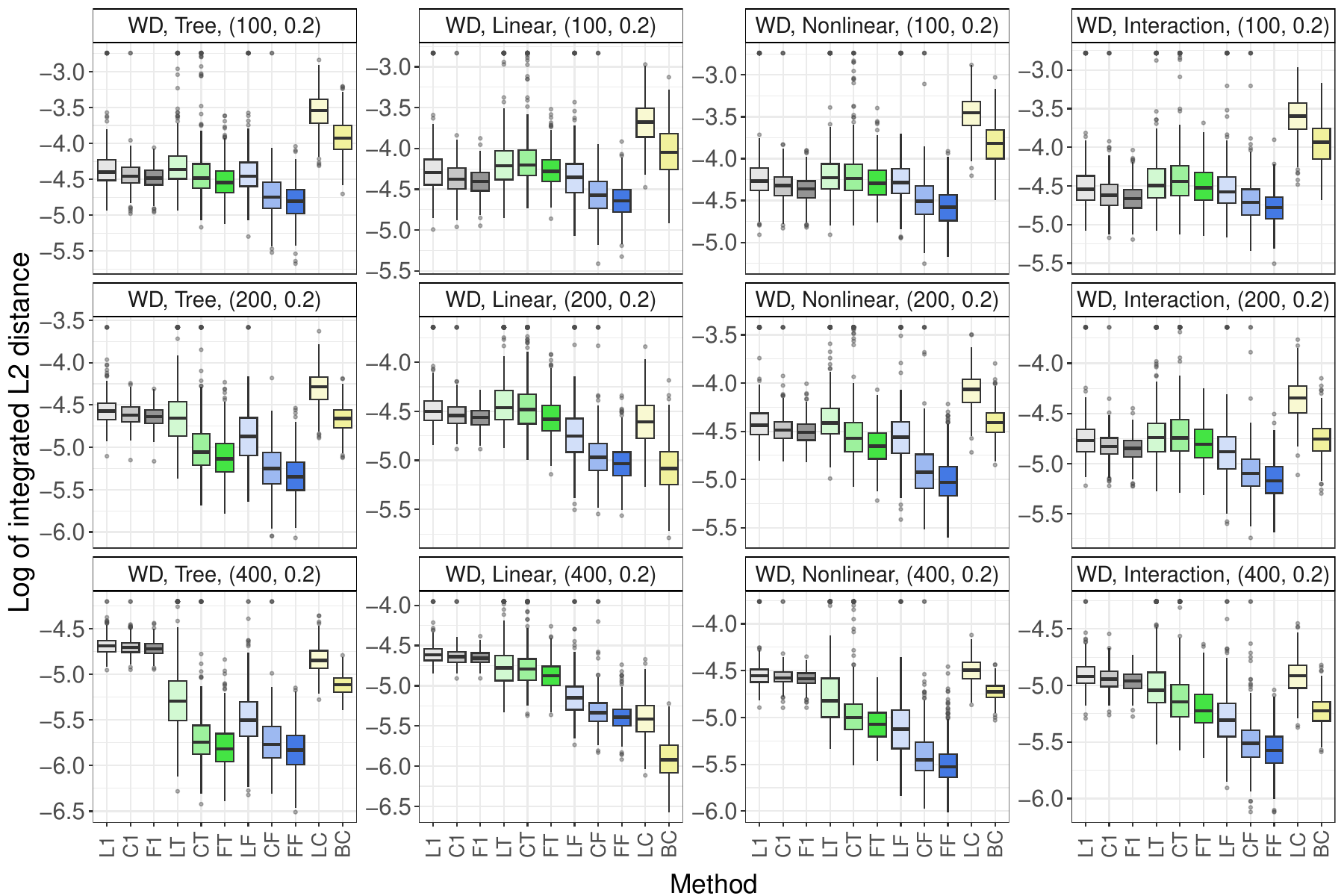}
    \caption{Log-scale integrated $L_2$-distance under Weibull decreasing (WD) hazard with 20\% censoring, for the four structures (tree, linear, nonlinear, interaction) and $n=100,200,400$; Panel labels denote $(n,c)$. Grey, green, blue, and yellow boxes denote one-sample estimators, CITs, CIFs, and Cox models, respectively, with lighter to darker shades corresponding to LTRC, MCLE, and MFLE variants within each method class. Extreme outliers beyond the upper whisker are capped.}
    \label{figS:prediction_WD_20}
\end{figure}

\begin{figure}[!h]
    \centering
    \includegraphics[width=\textwidth]{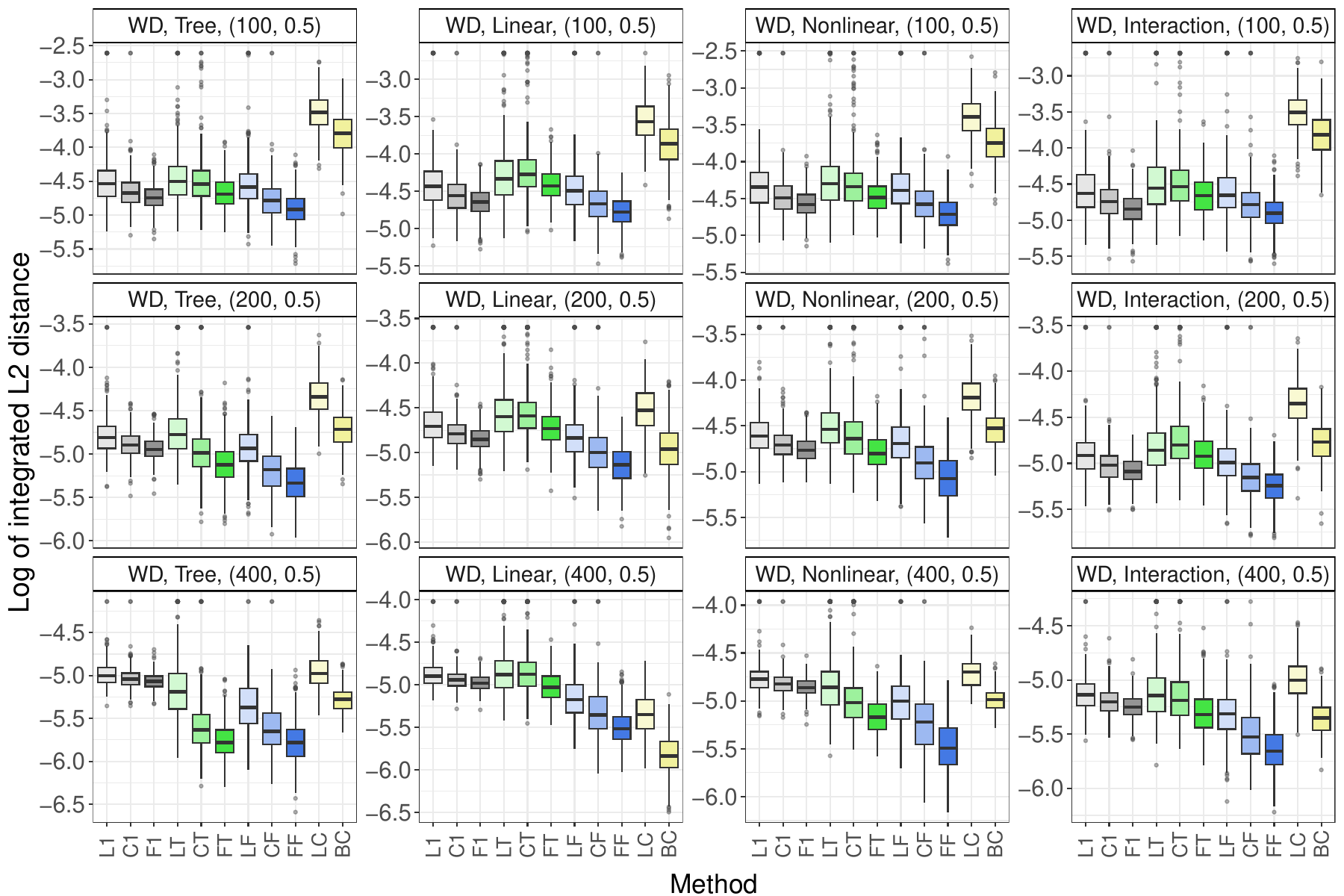}
    \caption{Log-scale integrated $L_2$-distance under Weibull decreasing (WD) hazard with 50\% censoring, for the four structures (tree, linear, nonlinear, interaction) and $n=100,200,400$; Panel labels denote $(n,c)$. Grey, green, blue, and yellow boxes denote one-sample estimators, CITs, CIFs, and Cox models, respectively, with lighter to darker shades corresponding to LTRC, MCLE, and MFLE variants within each method class. Extreme outliers beyond the upper whisker are capped.}
    \label{figS:prediction_WD_50}
\end{figure}

\begin{figure}[!h]
    \centering
    \includegraphics[width=\textwidth]{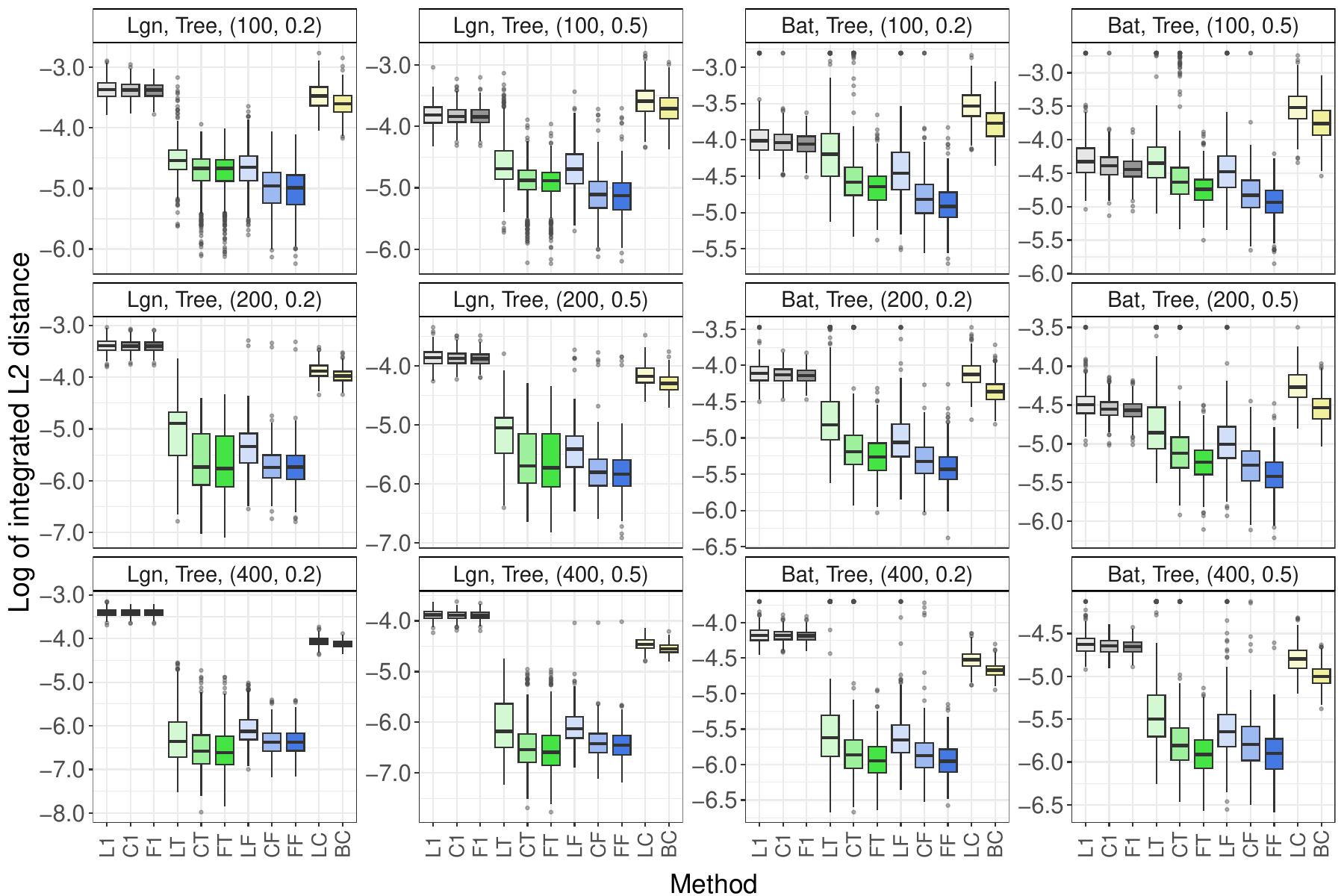}
    \caption{Log-scale integrated $L_2$-distance under log-normal (Lgn) and bathtub-shaped (Bat) hazards with 20\% and 50\% censoring under tree-structured settings and $n=100,200,400$; Panel labels denote $(n,c)$. Grey, green, blue, and yellow boxes denote one-sample estimators, CITs, CIFs, and Cox models, respectively, with lighter to darker shades corresponding to LTRC, MCLE, and MFLE variants within each method class. Extreme outliers beyond the upper whisker are capped.}
    \label{figS:prediction_LgnBat}
\end{figure}

Figures~\ref{figS:prediction_WI_50}--\ref{figS:prediction_LgnBat} present additional prediction results under Weibull increasing hazards with 50\% censoring, Weibull decreasing hazards with 20\% and 50\% censoring, and log-normal and bathtub-shaped hazard settings. Overall, the qualitative patterns are broadly consistent with those presented in the main manuscript. Across most settings, the proposed LBRC-based methods generally achieve lower prediction error than their corresponding LTRC-based counterparts, with the improvement becoming more pronounced under heavier censoring. Forest methods also tend to outperform single-tree methods, particularly in nonlinear and interaction settings. 

Differences between MCLE- and MFLE-based methods remain generally modest, although MFLE-based variants occasionally exhibit slightly lower prediction error under heavier censoring or Weibull decreasing hazards. Similar trends are observed across the additional log-normal and bathtub-shaped hazard settings, although variability is somewhat greater in smaller-sample scenarios.

under log-normal (Lgn) and bathtub-shaped (Bat) hazards with 20\% and 50\% censoring under tree-structured settings 

\clearpage

\section{Sensitivity analysis under misspecified truncation mechanisms}

The proposed LBRC-CIT and LBRC-CIF methods assume stationarity of the onset process, under which the truncation time follows a uniform distribution conditional on survival. To investigate robustness against violations of this assumption, we conducted sensitivity analyses under two forms of truncation misspecification.

Throughout the sensitivity analyses, the unbiased failure time $\tilde{T}$ followed a Weibull distribution with increasing hazard, with shape parameter $\alpha = 2$. The censoring rate and sample size were fixed at $c = 0.2$ and $n = 200$, respectively.

\begin{figure}[!h]
    \centering
    \includegraphics[width=0.6\textwidth]{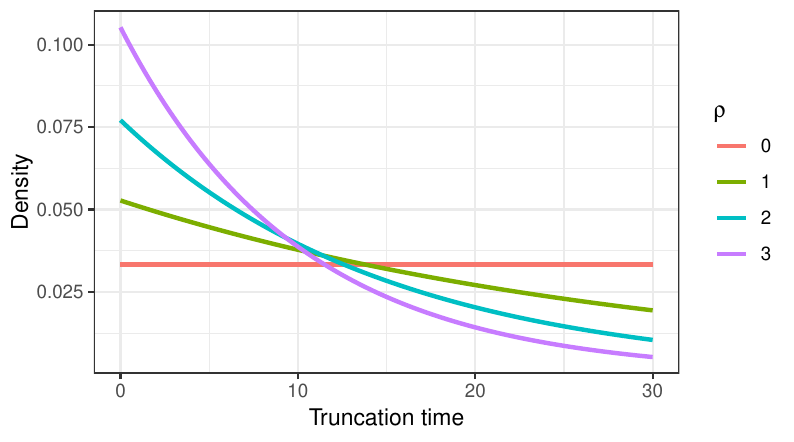}
    \caption{Density curves of the tilted truncated-exponential distribution for varying \(\rho\)}
    \label{figS:Texp}
\end{figure}

In the first scenario, all observations shared the same non-uniform truncation distribution (Texp). Specifically, the truncation time $\tilde{A}$ was generated from the tilted truncated-exponential distribution
\[
f(a;\rho,\tau)=
\begin{cases}
\dfrac{\rho/\tau \exp(-\rho a/\tau)}{1-\exp(-\rho)}, & \rho>0, \\
1/\tau, & \rho=0,
\end{cases}
\qquad 0 \le a \le \tau,
\]
where $\tau$ was chosen larger than the upper bound of $\tilde{T}$ to ensure length-biased sampling when $\rho=0$. As $\rho$ increases, the truncation distribution progressively deviates from the uniform distribution implied by stationarity and approaches an exponential-shaped distribution (Figure~\ref{figS:Texp}).

In the second scenario, the truncation mechanism depended on covariate values (Covd):
\[
\tilde{A}\mid X_1 \sim I(X_1 \ge 3)U(0,\tau)+I(X_1<3)f(a;\rho,\tau),
\]
thereby inducing differential truncation distributions across covariate-defined subgroups.

We evaluated three aspects of performance under these misspecified truncation distributions:
(i) unbiasedness of variable selection,
(ii) recovery of the true tree structure, and
(iii) survival prediction accuracy. Throughout the sensitivity analyses, simulation results are summarized using \(\mu=\rho/\tau\), a \(\tau\)-scale-invariant measure of departure from the uniform truncation distribution, where larger values indicate stronger deviations from the stationarity assumption.

\clearpage
\subsection{Unbiasedness of variable selection}
To evaluate unbiasedness of variable selection, we generated null data similarly to Section B of the Supplementary Material. The unbiased failure time $\tilde{T}$ followed a Weibull distribution with increasing hazard and scale parameter $\beta = 3$. The observed response for each observation was the triplet $(Z,A,\delta)$, where $Z=\min(T,A+C)$. Six independent covariates $X_1,\ldots,X_6$ were generated, where $X_1,X_2$ randomly take values from the set $\{1,2,3,4,5,6\}$, are $U[0,1]$, $X_3,X_4$ are binary$\{0,1\}$, and $X_5,X_6$ are $U[0,1]$.

\begin{figure}[!h]
    \centering
    \includegraphics[width=\textwidth]{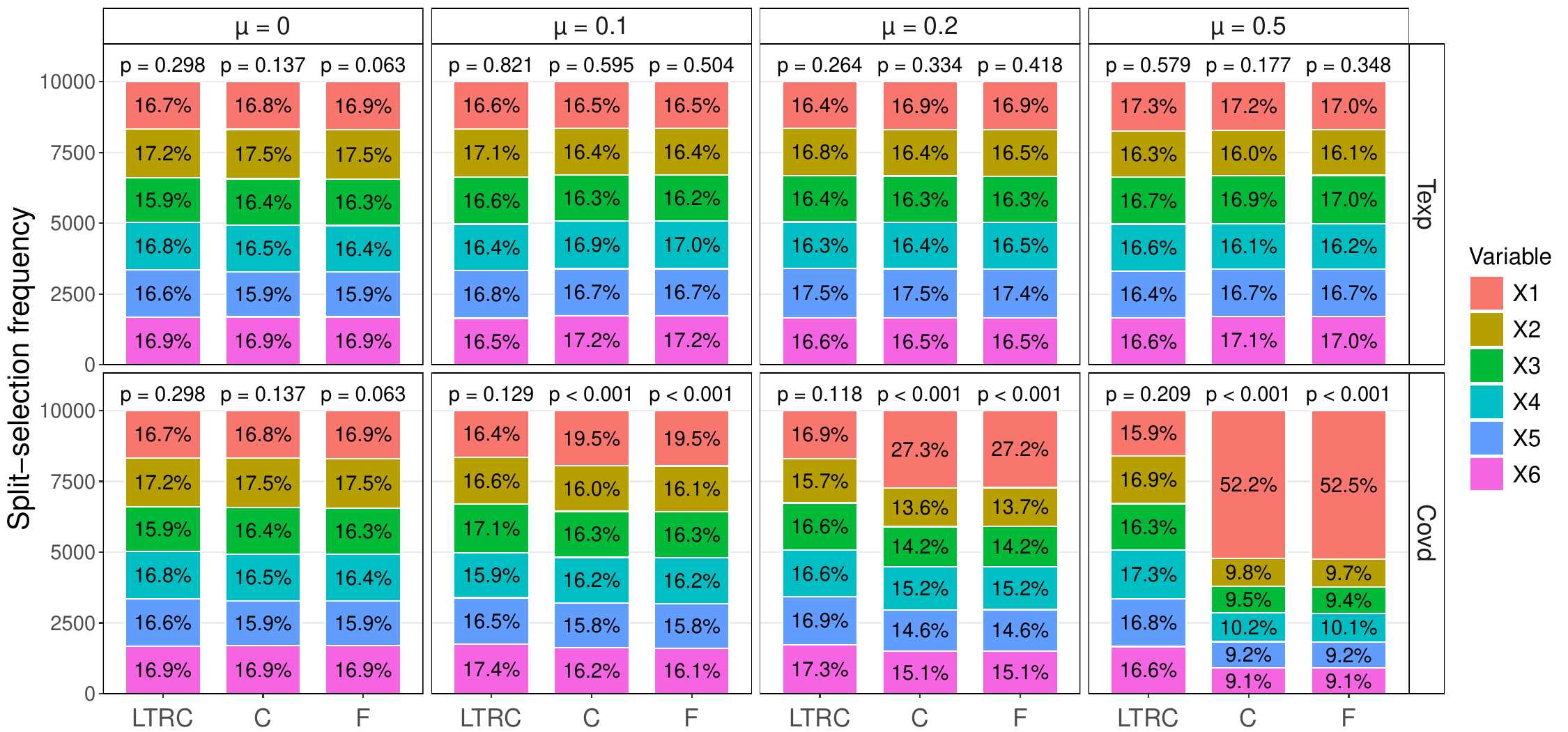}
    \caption{Root split-selection frequencies under truncation misspecification settings. LTRC, C, and F denote LTRC-CIT, LBRC-CIT(C,\(\cdot\)), and LBRC-CIT(F,\(\cdot\)), respectively. Percentages within bars are based on 10{,}000 simulation replicates.}
    \label{figS:Unbias_Sens}
\end{figure}

Figure~\ref{figS:Unbias_Sens} displays the root split-selection frequencies under the two truncation misspecification scenarios. Under homogeneous truncation misspecification (Texp), all three methods maintained approximately uniform split-selection frequencies across covariates for all values of \(\mu\). In contrast, under covariate-dependent truncation (Covd), the split-selection frequencies of LBRC-CIT-C and LBRC-CIT-F became increasingly distorted as \(\mu\) increased, indicating loss of unbiased variable selection. In particular, the proposed LBRC-CIT methods increasingly favored the covariate inducing the covariate-dependent truncation mechanism.

This behavior is consistent with the robustness patterns reported by Ning, Qin, and Shen (2010), who showed that score tests for length-biased data retained valid type-I error when truncation mechanisms were similarly misspecified across groups, but exhibited inflated type-I error when truncation distributions differed between groups. Since conditional inference trees repeatedly partition observations into covariate-defined subgroups and apply permutation-based split tests, a similar phenomenon is expected under covariate-dependent truncation mechanisms.

\clearpage
\subsection{Tree recovery and prediction accuracy}
We next evaluated tree recovery and prediction accuracy under the same tree-structured and nonlinear data-generating settings considered in the main simulation study. Both the Tree and Nonlinear settings were evaluated under the Texp and Covd truncation scenarios.

\begin{figure}[!h]
    \centering
    \includegraphics[width=\textwidth]{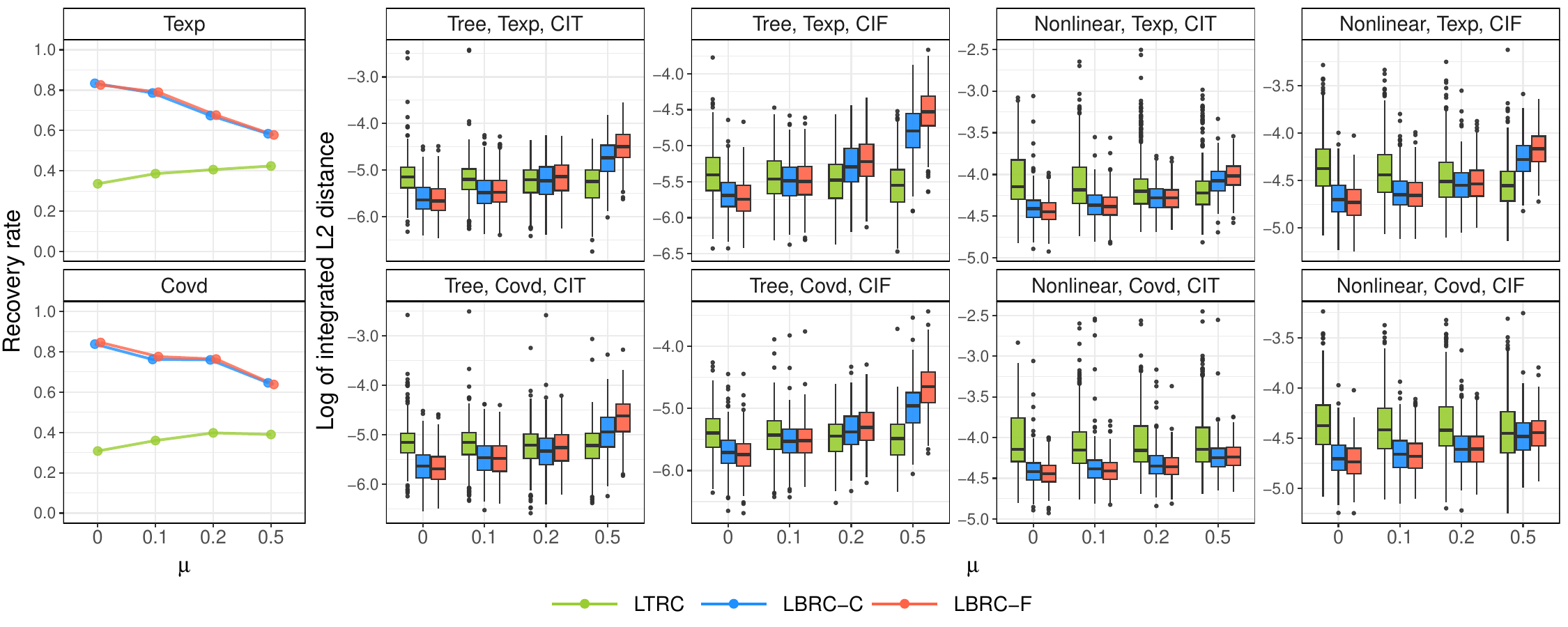}
    \caption{Recovery rates and prediction performance under Tree and Nonlinear settings across truncation misspecification levels.}
    \label{figS:RR_Pred_Sens}
\end{figure}

Figure~\ref{figS:RR_Pred_Sens} presents the recovery rates and prediction performance across varying levels of truncation misspecification. Although recovery rates gradually decreased as \(\mu\) increased, both LBRC-CIT variants consistently achieved higher recovery rates than LTRC-CIT across both truncation scenarios. The deterioration in recovery performance was also less pronounced under the Covd setting, where only a subset of observations departed from the length-biased assumption. This robustness pattern parallels the findings of Ning, Qin, and Shen (2010), where score tests for length-biased data remained more powerful than truncation-adjusted log-rank tests even under general left-truncated settings.

Prediction accuracy gradually deteriorated as the truncation distribution deviated further from the length-biased assumption. While the prediction performance of the proposed LBRC-CIT/CIF methods was comparable to or better than that of the LTRC methods under mild misspecification, larger prediction errors were observed for increasing values of \(\mu\). Although the proposed LBRC tree methods often continued to recover the underlying partitioning structure more accurately, prediction errors associated with truncation-model misspecification appeared to accumulate within terminal-node survival estimation.

\clearpage
\section{Detailed covariate definitions of real data example}
This section provides additional information on the dataset and  used in the analysis.

\subsection{Cancer Public Library Database}
Cancer Public Library Database (CPLD) is a comprehensive cancer data repository in South Korea, established in 2022 under the K-CURE project.
The CPLD integrates four major population-based sources: the Korea Central Cancer Registry (KCCR) for cancer incidence, Statistics Korea for death certification, and both the National Health Insurance Service (NHIS) and Health Insurance Review \& Assessment Service (HIRA) for health insurance eligibility, national health screenings, and healthcare service claims. 
The database includes patient demographics, medical history, cancer incidence and histologic subtypes, as well as long-term outcomes such as mortality and healthcare resource utilization, covering nearly 2 million individuals diagnosed with cancer.
Detailed information on CPLD is introduced in \textcite{choi2024data}.

\subsection{Variables used in eligibility criteria and covariate}
We collected variables including age, sex, body mass index, physical activity, smoking status, comorbidities, blood examination results, histologic subtype, SEER stage at diagnosis, and initial treatment. Specifically,

\begin{itemize}
    \item Body mass index is categorized as $<$18.5, 18.5–24.9, and $\geq$25. 
    \item Moderate physical activity refers to activities that increase breathing and heart rate, often resulting in sweating. 
    Vigorous physical activity is characterized by rapid breathing, to the extent that speaking more than a few words without pausing for breath becomes difficult. 
    \item A never smoker is defined as an individual who has smoked fewer than 100 cigarettes in their lifetime.
    \item Respiratory comorbidities include chronic bronchitis, bronchiectasis, and interstitial lung disease. 
    Other medical comorbidities include hypertension, diabetes mellitus, dyslipidemia, myocardial infarction, congestive heart failure, cerebrovascular disease, hemiplegia, chronic liver disease, chronic kidney disease, other malignancy than lung cancer, leukemia, lymphoma, connective tissue disease, peripheral vascular disease, and peptic ulcer disease.
    \item Blood examination results include hemoglobin, fasting glucose, aspartate transaminase, alanine transaminase, creatinine, and eGFR.
    \item Histologic subtypes of lung cancer are classified into squamous cell carcinoma, adenocarcinoma, large cell carcinoma, small cell carcinoma, non-specified carcinoma, and others. These are classified according to the International Classification of Diseases for Oncology, 3rd Edition (ICD-O-3) codes, consistent with the methodology of a previous studies (\cite{shin2017lung}, \cite{lee2025copd}).
    \item The stage of lung cancer are assessed according to the Surveillance, Epidemiology, and End Results (SEER) summary staging system (6). 
    The localized stage is defined as the cancer that confined to the lung, with no spread to other organs. 
    The regional stage is defined as the cancer that had spread to the chest wall, nearby organs, or regional lymph nodes. 
    The distant stage is defined as the cancer that had metastasized to distant organs, far from the primary lung cancer site.
    \item Initial treatment for lung cancer is defined as the treatment administered within the first four months following the diagnosis. 
    The treatments are categorized into surgery, chemotherapy, and radiotherapy, with patients potentially receiving more than one type of treatment. 
    Chemotherapy includes cytotoxic chemotherapy, epidermal growth factor receptor (EGFR) tyrosine kinase inhibitor (TKI), anaplastic lymphoma kinase (ALK) TKI, and immunotherapy.
\end{itemize}
As our focus is on the survival of patients with clinically serious lung cancer who did not initiate treatment shortly after diagnosis, the eligible cohort was defined by the following criteria:
\begin{enumerate}
\item Age ≥ 65 years at diagnosis;
\item Distant stage at diagnosis, as defined by the SEER summary staging system;
\item No lung cancer treatment (surgery, radiotherapy, or chemotherapy) within four months of diagnosis.
\end{enumerate}
Accordingly, the statistical analysis included 34 covariates: sex, body mass index, moderate or vigorous physical activity, five blood test measures, smoking status, three respiratory comorbidities, fifteen other medical comorbidities, and six histologic subtypes.

\printbibliography

\end{refsection}

\end{document}